\documentclass[draft]{agujournal2019}
\usepackage{url} 
\usepackage{lineno}
\usepackage{physics}
\usepackage{amssymb}
\usepackage[inline]{trackchanges} 
\usepackage{soul}
\usepackage{import}
\usepackage{longtable}
\usepackage{lscape}
\usepackage{rotating}
\usepackage{bm}
\usepackage{pifont}
\usepackage{graphicx}

\newcommand{\xmark}{\text{\ding{55}}}

\journalname{JAMES}

\begin{document}

\title{A comparison of data-driven approaches to build low-dimensional ocean models}

\authors{N. Agarwal\affil{1}, D.Kondrashov\affil{3,5}, P. Dueben\affil{2}, E.Ryzhov\affil{1,4}, P. Berloff\affil{1}}

\affiliation{1}{Department of Mathematics, Imperial College London, London, SW7 2AZ, UK}
\affiliation{2}{ECMWF, Shinfield Road, Reading, RG2 9AX, UK}
\affiliation{3}{Department of Atmospheric and Oceanic Sciences, University of California, Los Angeles, CA 90095, USA}
\affiliation{4}{Pacific Oceanological Institute, Vladivostok, 690041, Russia}
\affiliation{5}{Institute of Applied Physics of the Russian Academy of Sciences, 603950, Nizhny Novgorod, Russia}

\correspondingauthor{Niraj Agarwal}{n.agarwal17@imperial.ac.uk}

\begin{keypoints}
\item The multi-level stochastic approach produces the most stable, accurate, and low-cost emulator of a double-gyre ocean model solution.
\item ANN and LSTM work better in a hybrid form with linear regression, providing the core dynamics, than in their standalone application.
\item Emulators incorporating memory effects and state-dependent noise show enhanced performance and deep learning can learn these effects. 
\end{keypoints}

\newpage
\begin{abstract}
We present a comprehensive inter-comparison of linear regression (LR), stochastic, and deep-learning approaches for reduced-order statistical emulation of ocean circulation.
The reference dataset is provided by an idealized, eddy-resolving, double-gyre ocean circulation model.
Our goal is to conduct a systematic and comprehensive assessment and comparison of skill, cost, and complexity of statistical models from the three methodological classes.

The model based on LR is considered as a baseline. 
Additionally, we investigate its additive white noise augmentation and a multi-level stochastic approach, deep-learning methods, hybrid frameworks (LR plus deep-learning), and simple stochastic extensions of deep-learning and hybrid methods.
The assessment metrics considered are: root mean squared error, anomaly cross-correlation, climatology, variance, frequency map, forecast horizon, and computational cost. 

We found that the multi-level linear stochastic approach performs the best for both short- and long-timescale forecasts.
The deep-learning hybrid models augmented by additive state-dependent white noise came second, while their deterministic counterparts failed to reproduce the characteristic frequencies in climate-range forecasts.
Pure deep learning implementations performed worse than LR and its noise augmentations.
Skills of LR and its white noise extension were similar on short timescales, but the latter performed better on long timescales, while LR-only outputs decay to zero for long simulations.

Overall, our analysis promotes multi-level LR stochastic models with memory effects, and hybrid models with linear dynamical core augmented by additive stochastic terms learned via deep learning, as a more practical, accurate, and cost-effective option for ocean emulation than pure deep-learning solutions.
\end{abstract} \newpage
\section*{Plain Language Summary}
In weather and climate predictions, scientists use comprehensive ocean circulation models for representing the effects of the oceans on the atmosphere. 
These models simulate the three-dimensional ocean dynamics using millions of variables and, thus, require significant computational resources and running time. 
Therefore, there is a need for low-cost, data-driven ocean models with fewer variables that can reproduce essential oceanic circulations with reasonable accuracy.
There are several popular data-driven approaches to build these models, but singling out the best one is difficult and significantly understudied.
We have systematically assessed and compared the accuracy, stability, and computational cost of various data-driven models against the linear regression -- a fundamental and easy-to-implement deterministic model, i.e., it provides a fixed output for a fixed input.
We considered several stochastic and deep-learning models for comparison; stochastic models combine a deterministic model with customized noise, whereas deep-learning models train a complex network of neurons similar to the human brain.
We found that the stochastic models that properly include the core dynamics, time-delay effects, and model errors perform the best. 
The core dynamics provides the essential changes, time-delay effects are the changes due to correlation between successive ocean states, and model errors provide other possible causes of changes.  
\newpage
\section{Introduction}
\label{sec:introduction}
Medium-range weather forecast models routinely use computationally expensive Ocean General Circulation Models (OGCMs) that are coupled to the atmosphere model.
However, the long timescales of ocean dynamics and the weak influence from the deeper layers of the ocean on the atmosphere for a weather forecast of, say, a couple of days justify the investigation of replacements of expensive OGCMs with low-dimensional data-driven models that can run at negligible cost and emulate the upper ocean. 
Here, these models are referred to as ocean ``emulators'', because they emulate statistical properties of the flow rather than simulating the dynamics derived from physical principles. 
Our definition of emulators is slightly different when compared to a number of studies that try to replace components of existing models to reduce computational cost. 
In our case, we do not aim to emulate a model component but rather the physical system that has generated the data -- a dynamical ocean model.
Ocean emulators are also useful (i) in long-time climate-type model simulations for process sensitivity studies, (ii) in climate prediction, and (iii) for improving ensemble forecast statistics.
Furthermore, ocean emulators can potentially be down-scaled and used for data-driven parameterizations of mesoscale (and even sub-mesoscale) eddies for non-eddy-resolving and eddy-permitting comprehensive OGCMs.
Finally, they can also be used as conceptual toy models for process-related studies (e.g., as kinematic flow emulator of material transport).

The physical ocean models have the fundamental advantage that they can operate even without training from data, which simply may not exist.
The main problems with the physical models are that for some practical applications they can be prohibitively expensive or may not allow to resolve all important features, some of the involved physics can be inaccurately accounted for, and numerical and discretization errors can be unacceptable.
On the other hand, data-driven emulators, which are the focus of this study, can be much cheaper, accurate (for the data-trained regimes), and simple to deal with, which gives them a crucial advantage for many practical problems.
However, they can be hard to interpret physically as many of them are ultimately used as a black box, e.g., machine-learning-based methods.
Low-cost emulators can be constructed in terms of the Empirical Orthogonal Functions (EOFs) and their Principal Components (PCs; \citeA{lorenz1956empirical}), but the true governing equations in the EOF space are always unknown.
Our approach to ocean emulation is supported by the existing and rapidly developing methodologies for statistical data-driven modeling (reviewed in \citeA{rowley2017model, brenner2019perspective}).
We consider three major statistical model types: linear regression (LR), stochastic, and deep learning.

LR belongs to the broader family of regression-only statistical models, where some polynomial surfaces are fit to the data, while forbidding correlations with the residuals.
The fitted polynomials are believed to capture ``enough'' of the dynamics, so that the residuals can be attributed to the uncertainty in the initial conditions and internal variability.
The regression-only models benefit from significant speed-up as there is no need to identify the covariance matrix and the associated calculations, such as covariance inversion, which can be complex and computationally expensive for high-dimensional systems. 
The fast, simple, and easy-to-implement characteristics of regression-only models found them numerous applications in climate and environmental sciences \cite{sexton2012multivariate, holden2013plasim, williamson2015exploratory}.
Here, we use them in the linear form, due to incomplete knowledge about the nonlinear basis functions of the ocean circulation tendencies, and develop the simplest ocean emulator that provides the baseline. 

Next, stochastic models use parameterized noise signals to deal with missing/unknown physics, parameter uncertainties, inaccurate initial conditions, and noise-induced regime transitions \cite{sardeshmukh2001rossby, sura2005multiplicative, berner2017stochastic} in chaotic dynamical systems.
They produce random outputs, thus, allowing for ensemble statistics for uncertainty estimates.
Stochastic terms can be added to any deterministic framework in two common ways, as either additive or multiplicative noises.
The additive noise is directly added to the basic equations, whereas the multiplicative noise is added after multiplying it with the amplitude function depending on the predicted model variables.
The latter models are significantly more complicated, and fitting their parameters is more difficult, yet both have distinct advantages and are applied in ocean and climate modelling for numerous purposes.
The additive noise has been used to force linear dynamical models \cite{farrell1993stochastic, delsole1999empirical, zhang1999linear}, to model effects of subgrid-scale turbulence \cite{farrell1995stochastic, d2001extratropical, berloff2003material, delsole2004stochastic, williams2016improved}, to provide stochastic climate predictions \cite{majda1999models, seiffert2008impact, seiffert2010stochastic}, and to derive stochastic primitive equations for the oceans and atmosphere \cite{ewald2007stochastic}.
On the other hand, the multiplicative noise strategy is considered most relevant for modelling non-Gaussian statistics, such as extreme events, tipping points in the dynamical systems \cite{sura2011general, franzke2012predictability, franzke2013predictions, sura2013stochastic}, uncertainty estimates in parameterization schemes \cite{buizza1999stochastic, juricke2013effects, juricke2017stochastic, ollinaho2017towards}, stochastic primitive equations \cite{glatt2008stochastic, debussche2012global}, and nonlinear coupling between noise and model variables \cite{sardeshmukh2003drifts, franzke2005low}.

Recently, a multi-level framework of the additive nonlinear stochastic models -- called Empirical Model Reduction \cite[EMR]{kondrashov2005a,KravtsovKondrashovGhil_JCL05,KravtsovGhilKondrashov_09,KCRG13} -- has also been developed.
It allows to include inherently important time-delayed (memory) effects both in the additive  state-dependent stochastic forcing and dynamical operator \cite{MSM2015},  and allows for a more complicated temporal structure of the noise. Data-driven climate models based on EMR formulation have proven to be highly competitive in prediction and process studies \cite{Strounine2010,Chen_etal2016,GGKR18}. 
The Linear inverse modeling (LIM) approach \cite{Penland_MWR89, PenlandSardeshmukh_JCL95, Penland_PD96} is a specific case of the EMR  with a linear propagator and additive white noise. 
\citeA{kondrashov2015gyres} have shown that decadal oceanic variability can be successfully simulated by a linear EMR formulation and a change of basis, namely instead of Principal Component Analysis (PCA) modes,  using modes identified by Multichannel Singular Spectrum Analysis (M-SSA) which incorporates time-delayed embedding \cite{Ghil.review.ea.2002}. 
Similar to M-SSA, data-adaptive harmonic decomposition (DAHD)  \cite{kondrashov2018multiscale,KRB2020,RKNB2020} relies on the eigendecomposition of the lag-covariance matrix. However, unlike M-SSA, DAHD modes form an orthonormal set of spatial patterns oscillating harmonically within the time-embedding window, and thus can be modeled by a system of coupled frequency-ranked nonlinear stochastic oscillators. 

For this study in the class of stochastic methods, in order to make a fair comparison, we have made a deliberate choice to focus on the methods that are commonly used in PCA basis, and implemented linear stochastic models for white noise, as well as the linear formulation of the multi-level EMR framework. 

The third statistical model type investigated in this study are deep-learning models. 
These models approximate the intricate nonlinear functional relationships between the model inputs and outputs by training an extensive parametric network of interconnected nodes, using neither physical knowledge about the system nor the governing differential equations. 
With the recent advancements in computing power, simple feed-forward Artificial Neural Networks (ANNs), Convolutional Neural Networks (CNNs), and Long Short Term Memory (LSTM) models rose to prominence in many science disciplines and helped to find hidden patterns in multi-dimensional data sets.
Feed-forward ANN is the most commonly used deep-learning model, and its design includes multiple layers of small blocks of equations communicating in a nonlinear fashion.
These ANNs have been used broadly in oceanic and atmospheric studies, ranging from the idealized Lorenz63 \cite{lorenz1963deterministic} and Lorenz96 \cite{lorenz1996predictability} models \cite{gmd-11-3999-2018, scher2019generalization} to more realistic situations, such as developing subgrid-scale models \cite{karunasinghe2006chaotic, rasp2018deep, maulik2019subgrid}, learning the inter-dependency between global climate and vegetation fields \cite{holden2015emulation}, super-parameterizations \cite{chattopadhyay2020data}, and spotting extreme events in complex weather data sets \cite{liu2016application}.
CNNs \cite{krizhevsky2012imagenet} form another class of deep-learning methods that has been extensively used in geophysical fluid dynamics to identify (and regress) patterns in turbulent flow regimes by repeatedly convoluting the inputs with appropriate kernels \cite{bolton2019applications,liu2020deep,weyn2020improving,chattopadhyay2020deep}.
LSTMs \cite{hochreiter1997long} and Reservoir Computing \cite{jaeger2004harnessing,RC,Nadiga2021} are specific forms of Recurrent NNs that can progress learned information from one timestep to the next when applied in an iterative way.  
This improves the time evolution of the model which makes the LSTMs attractive for applications in oceanic and atmosphere modelling that show multi-scale and lagged interactions \cite{zhang2017prediction, vlachas2018data, salman2018weather}.

Despite first attempts to interpret deep-learning models physically \cite{mcgovern2019making, portwood2021interpreting}, they remain mostly black boxes.
Therefore, the modern trend is to use them in combination with physical models \cite{karpatne2017theory, reichstein2019deep} --- referred to as hybrid methods --- to potentially increase the forecast skills of an imperfect physical model.
Also, the involved neural network may require less training and complexity \cite{jia2019physics}.
A few applications of hybrid methods using ANN and LSTMs are in \citeA{krasnopolsky2006complex, rahman2018hybrid,watson2019applying,pawar2020data}.
Similarly, hybrid deep-learning stochastic approaches are also being developed \cite{mukhin2015,Seleznev2019}, but this area remains understudied, and its full potential has not yet been explored. 
In this paper, we test all of the above deep-learning models, except for the CNN, which is best suited to image-based datasets and structured grids, whereas, here, we work in the EOF/PC space to achieve significant model order reduction. 
We also propose novel hybrid stochastic formulations by utilizing residuals from the deep-learning procedure, thus effectively providing nonlinear state-dependent noise. 

How do different emulators from the selected 3 classes compare against each other in terms of their skills?
This is a difficult and significantly understudied question, which is central in our study. In particular, as most papers will only evaluate a single method for a specific dataset, quantifiable intercomparison of different methods are often impossible across papers.
In the context of geophysical applications, we can speculate the LR to be the least successful due to its purely linear form and deterministic nature, making it ineffective in accounting for the inherent uncertainties due to, for e.g., insufficient resolution, unresolved processes, parameterization errors, etc., and imprecise/incomplete knowledge of many geophysical processes (especially on the reduced-dimension space) and scale interactions caused by non-linear terms of the differential equations.
The stochastic models can deal with model and forecast uncertainties and, therefore, are expected to perform better than LR.
However, it is hard to predict their performance relative to the deep-learning methods, which are generally deterministic (note that they can also be Bayesian) but capable of producing accurate and generalizable models.
In this paper, we aim to compare a number of ocean emulators for relatively simple ocean circulation data obtained from a long-term model simulation of an idealized ocean model. We do not use ocean observations as model data to avoid problems due to measurement errors, complex coast-lines, gaps in observations, biases between different observational products, etc. which would make a fair comparison between the different emulators more difficult. However, the study could easily be repeated with observational datasets.
We aim to develop emulators, which are computationally cheap, able to reproduce the statistical characteristics of the reference flow, and are capable of providing simulations on climate-range time scales.

Section \ref{sec:dataset} provides details about the datasets; Section \ref{sec:models} discusses all the modelling frameworks;  Section \ref{sec:results} presents the model assessment metrics and their outcomes, and Section \ref{sec:conclusions} discusses the results and concludes. 
\section{Dataset}
\label{sec:dataset}
\begin{figure}
 \includegraphics[width=\textwidth]{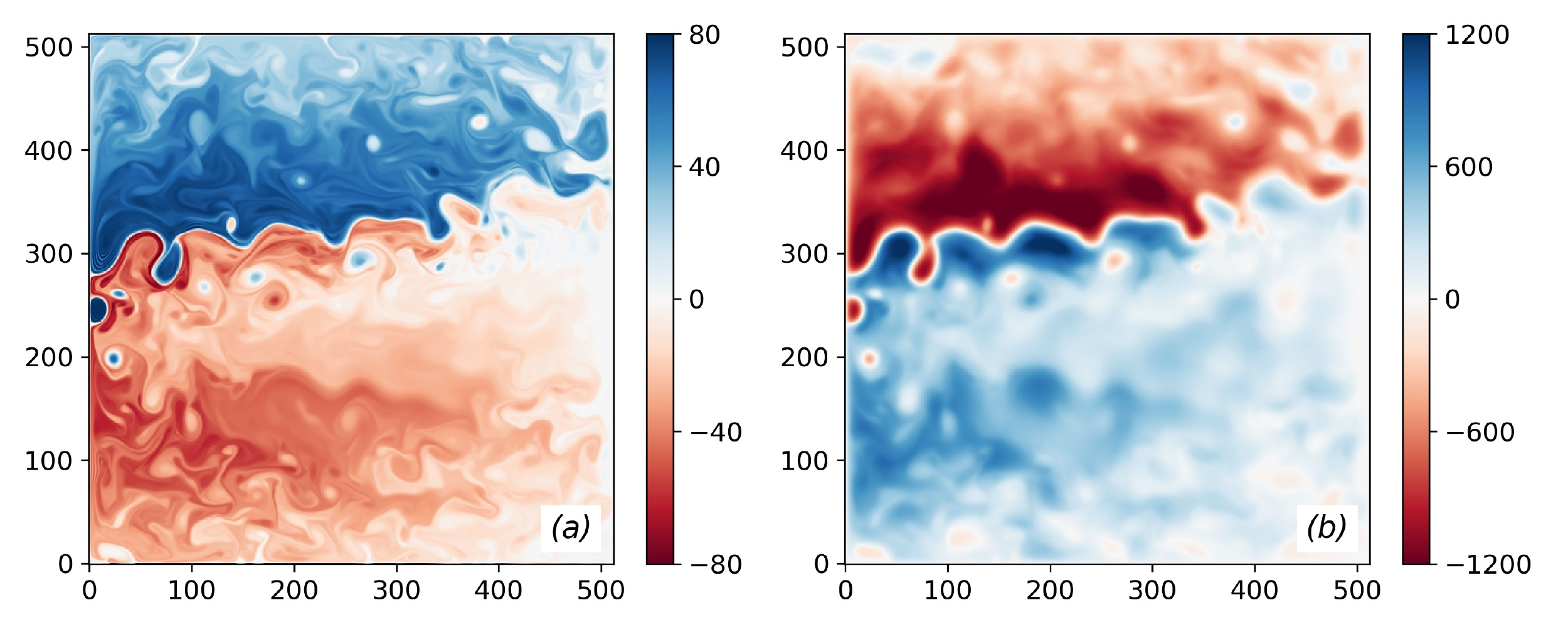}
 \caption{
 Illustration of the reference eddy-resolving ocean circulation solution.
 Snapshots of the top-layer (a) potential vorticity anomaly and (b) velocity streamfunction. 
 In case of (a), the anomaly is computed by subtracting the planetary vorticity from the full potential vorticity.
 Positive (red) and negative (blue) values of potential vorticity anomaly correspond to cyclonic and anticyclonic motions, respectively, and oppositely for the streamfunction.
 The color scales are in dimensionless units.
 Non-dimensionalization is done using the length and velocity scales equal to $7.5$km (grid interval) and $0.01$ m/s, respectively, and CGS units.
 }
 \label{fig:flow_snapshot}
\end{figure}

The reference data set used in this study was generated by a three-layer double-gyre quasigeostrophic ocean model \cite{berloff2015dynamically, ryzhov2019data}, which provides an idealized representation of the North Atlantic wind-driven circulation dominated by the subtropical and subpolar gyres, and by the Gulf Stream current.
A square ocean basin with side $3840$ km is considered; a steady asymmetric wind forcing at the top layer is imposed; the partial-slip boundary condition is used on the lateral boundaries; stratification is imposed with the first and second Rossby deformation radii equal to $40$ km and $20.6$ km, respectively; and the grid resolution is $7.5$ km.
Since the model is dynamically eddy-resolving, it qualitatively correctly reproduces the eastward jet extension of the western boundary currents and its adjacent recirculation zones, as well as the mesoscale eddy variability and interdecadal oscillations (of period $17$ years).
The reference solution of the statistically stationary flow regime is obtained in terms of the evolving potential vorticity and velocity streamfunction that are related to each other via elliptical inversion.
The solution snapshots are saved after every 10 days, for a total of about 1400 years (5$\times 10^5$ days).
Both potential vorticity anomaly and streamfunction snapshots (Fig. \ref{fig:flow_snapshot}) show two asymmetric gyres of opposite circulations separated by the eastward jet region, which is characterized by the most vigorous flow variability, and, therefore, is in the focus of our study.

\begin{figure}
    \centering
    \includegraphics[width=\textwidth]{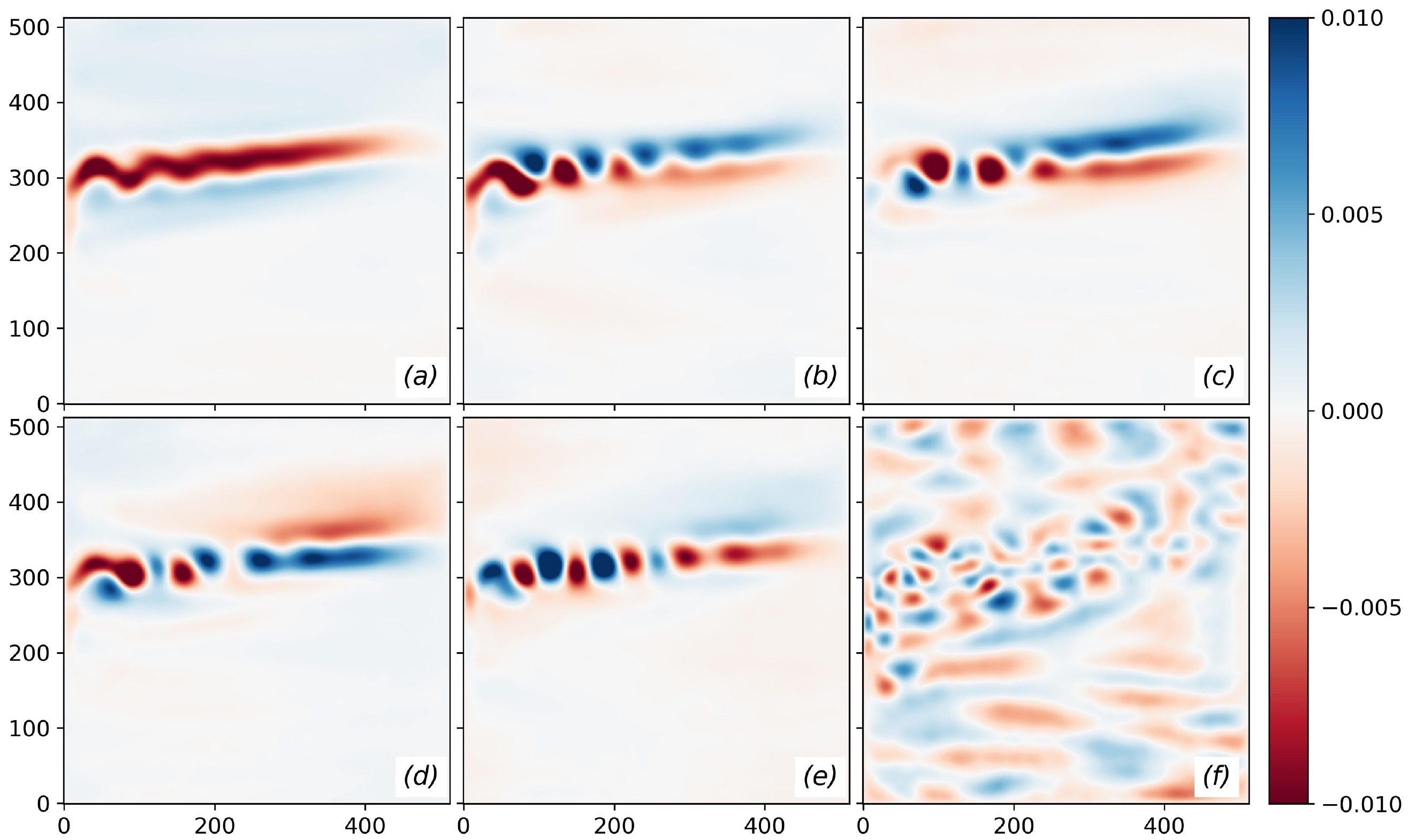}
    \caption{
    Representation of the reference data in terms of the EOFs of the upper-layer streamfunction.
    (a)-(e) represent the top-five and (f) represent the bottom-most EOFs from the selected suite of top-150 EOFs/PCs.  
    Note that all top ranked patterns significantly contribute to the multi-scale variability in the eastward jet region, which is the focus of our study, but the rest of the basin is also impacted.
    The bottom ranked EOF mostly accounts for small-scale variabilities in the entire domain. 
    The color scale values are in the dimensionless units; non-dimensionalization is done using the same scales as mentioned in Fig. \ref{fig:flow_snapshot}.
    }
    \label{fig:psi1_EOFs}
\end{figure}

For the reference data set, we chose velocity streamfunction, because it is smoother than potential vorticity anomaly (both of them represent the same dynamical regime).
We deal with the upper isopycnal layer, as it contains the most intensive flow variability and is the most relevant for numerical weather predictions.
Next, we re-organize the solution description and reduce the dimensionality of the raw data by using singular value decomposition (SVD), which allows us to find dominant spatial patterns, called Empirical Orthogonal Functions (EOFs), and the corresponding temporal coefficients, called Principal Components (PCs).
Before SVD, the reference 3D spatio-temporal dataset ($\in \mathbb{R}^{m\times m \times n}; m = 513, n = 5\times10^4$ with the grid-dimension in each direction $m$ and the number of records $n$) is reshaped into a matrix $(\in \mathbb{R}^{n \times m^2})$, where each row is a snapshot of the flow. 
It is then smoothed along the temporal dimension with a $100$-days (i.e., $10$ records) running-average window, in order to focus on the long-timescale tendencies (the window size is truncated at the endpoints), followed by an SVD decomposition to obtain $n$ EOFs/PCs.
These EOFs/PCs are used for the emulation and analyses in this work.
Each EOF explains a fraction of the total temporal variance, and all EOFs are ranked so that this fraction decreases with $k$; for complex geophysical data, this decreases exponentially \cite{Ghil.review.ea.2002}. 
For the purposes of our study, we considered the leading $150$ EOFs and their PCs, that all together represent about $90\%$ of the total variance. 
Our choice of relatively small ($0.3\%$ of the total number of EOFs), yet dynamically significant, number of EOFs is justified by our goal to gain a foothold in developing and applying a systematic methodology for comprehensive assessment of the model skills.
However, we admit that we are trying to build a simplified statistical model of the QG dynamics, which itself is a simplification of the comprehensive general circulation model dynamics, therefore, our study does not present the real ocean situation, and the generalization of the results presented here are only possible up to a certain extent.   
Fig. \ref{fig:psi1_EOFs} shows the top-five and the bottom-most EOFs among the $150$ EOFs/PCs considered here.
The top-ranked EOFs represent the most dominant patterns along the eastward jet region, which is in the focus of this study, as it contains most of the variability, while the last EOF represents high-frequency variabilities across the basin.

\begin{figure}
    \centering
    \includegraphics[width=\textwidth]{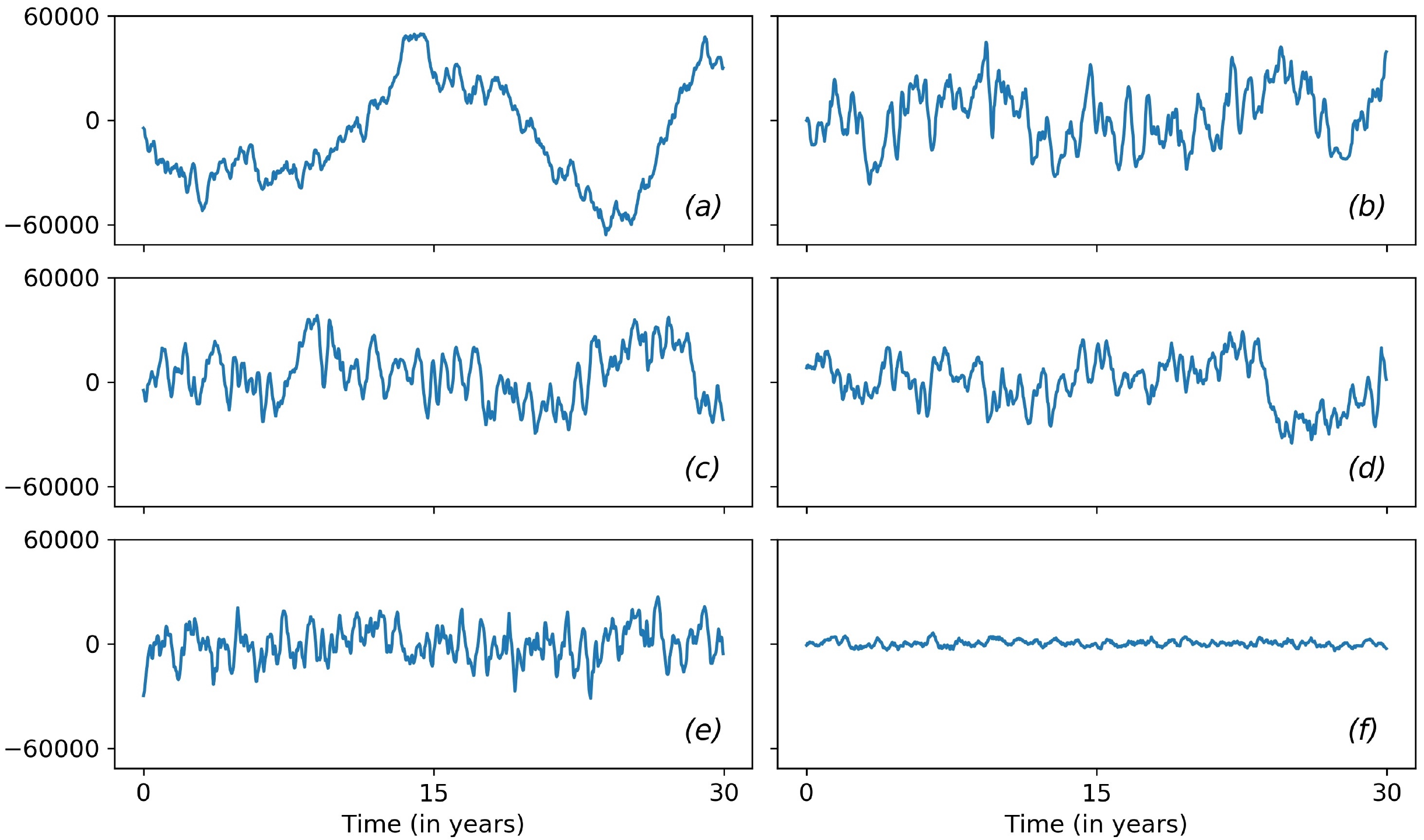}
    \caption{
    Representation of the reference data in terms of the PCs corresponding to the EOFs in Fig. \ref{fig:psi1_EOFs}.
    Note that PCs are shown only for $30$ years of the total record of about $1400$ years.
    Also, note that the low-ranked PC is much noisier with a smaller variance.
    The $y$-axis values are dimensionless.
    }
    \label{fig:psi1_PCs}
\end{figure}

Our aim is to emulate the leading $150$ PCs of the upper-ocean streamfunction (Fig. \ref{fig:psi1_PCs}) efficiently and to construct the corresponding reduced-order models.  
\section{Modelling Frameworks}
\label{sec:models}

In this section, we expand on each adopted modelling methodology one-by-one, by providing the mathematical formulations and the relevant parameters and hyper-parameters, and by explaining the model training processes.
Before modeling, we normalized the PCs by the respective standard deviations such that each of them follows zero mean and unit standard deviation. 
However, the results were found to be robust to other normalizations too, e.g., division by the standard deviation of the top-most PC (similarly for auxiliary variables, if any).
Within each method, we model either tendencies or state of the PCs, denoted by $y_i;\hspace{0.2cm} i = 1,2,3, \ldots, N$, where $N = 150$ is the total number of PCs considered. 
The tendencies are computed numerically using the finite-difference method with $\Delta t = 10$ days as the time difference between any two successive records.
Out of the total $50K$ records (which covers a time interval of $500K$ days, as we have recorded the snapshots after every $10$ days) of the PCs, we set the initial $40K$ records as the training dataset, and the last $10K$ records as the test dataset.
This training length is roughly $400$ times the decorrelation time scale of the topmost PC ($\approx 2.7$ years) and includes nearly $65$ periods of the intrinsic low-frequency variability of period $17$ years.
Note that such a partitioning of training and test datasets may leave the end of the training dataset and the beginning of the test dataset correlated.
But this is inconsequential because the test dataset still possesses roughly $16$ cycles of the intrinsic low-frequency variability, and we have considered multiple initial conditions for forecasts and reported their average.

We considered the full length of the training dataset for all models but also experimented with different lengths ranging from $1K$ to $40K$ records as in real-world applications, such a large training dataset (over $1000$ years) may not be available. 
Our analysis showed that at least $10\%$ of the training data (corresponding to $\approx 7$ cycles of low-frequency variability) is needed for our models to have similar performance on the test dataset as reported for the full length of training dataset.
Transfer learning may help when training data-driven models for the real ocean with limited observations. 
In this approach, the models are trained first using the historical climate data, such as CMIP model outputs, before fine tuning using observations and reanalysis data \cite{ham2019deep,rasp2021data}.
\subsection{Linear Regression (LR)}
\label{subsec:LR}
This model is expressed as follows:
\begin{linenomath*}
\begin{equation}
    \frac{\dd \mathbf{y}}{\dd t} = \mathbf{Ly}; \, ,
    \label{eq:SLR}
\end{equation}
\end{linenomath*}
where $\mathbf{y} = [y_1\hspace{0.2cm}y_2\hspace{0.2cm} \ldots \hspace{0.2cm}y_N]$ is the multivariate PC, and $\mathbf{L}$ is an $N\times N$ matrix of regression coefficients, referred to as the system matrix.
LR is the simplest model considered in this work and is used as a baseline for assessing the performance of the others.
Linear dynamics is pertinent to the gyres and explains a significant proportion of the low-frequency variabilities in the top PCs.
Therefore, we also term LR as the `core dynamics'.

\subsection{LR with Additive White Noise (LR-AWN)}
\label{subsec:AddLR}
This is a stochastic model based on the LR model (\ref{eq:SLR}) but additionally equipped with the additive white noise. 
The model is given as:
\begin{linenomath*}
\begin{equation}
    \dd \mathbf{y} = \mathbf{Ly} \dd t + \mathbf{Q}\dd\mathbf{W}; \, ,
    \label{eq:SLR_with_white_noise}
\end{equation}
\end{linenomath*}
where $\mathbf{W} = [W_1\hspace{0.2cm} W_2\hspace{0.2cm} W_3 \ldots W_N]$ is a multivariate Wiener process with $d\mathbf{W} \sim \mathcal{N}(\mathbf{0},\bm{\sigma}^2)$, where $\mathbf{\bm\sigma}$ is the standard deviations of the LR residuals $\mathbf{r} = \dd \mathbf{y}/\dd t - \mathbf{Ly}$; $\mathbf{Q}$ is a lower triangular matrix -- obtained by Cholesky decomposition of the correlation between the residuals $\mathbf{r}$ at zero lag -- multiplied to $\dd \mathbf{W}$ in order to induce the required correlation between the noise components. 
Mathematically, $\mathbf{QQ^T = C}, $ where $\mathbf{C} = Corr ([r_1\hspace{0.2cm}r_2\hspace{0.2cm}r_3\hspace{0.2cm} \ldots \hspace{0.2cm}r_N])$ and $\mathbf{Q, C} \in \mathbb{R}^{N\times N}$.
We validated the output of this model against a Linear Inverse Model, which is more widespread, but found almost no difference between the two model outputs. 

\subsection{Multi-level Linear Regression (ML-LR)}
\label{subsec:MLLR}
We adopt a linear formulation of multi-level stochastic Empirical Model Reduction approach  \cite{KravtsovGhilKondrashov_09,kondrashov2015gyres,MSM2015}, where the regression residuals are not immediately replaced by some noise but instead are modeled by using a stack of levels.
The top-most level is similar to (\ref{eq:SLR}), and the lower levels regress the higher-level residuals as the additional (hidden) state variables, until the lowest-level residual degenerates into spatially uncorrelated white noise, i.e., their autocorrelation approaches zero (technical implementation of the stopping criterion is based on fraction of the explained variance by regression, see Appendix A in \citeA{MSM2015}).
The complete model can be expressed as:
\begin{linenomath*}
\begin{subequations}
\begin{eqnarray}
	\text{Level 1:} &\dd {\bf y} & = {\bf Ly}{\dd t} + {\bf r}(t){\dd t} \, , \\
	\text{Level 2:} &\dd {\bf r}& ={\bf M^{(1)}}[{\bf y}; {\bf r}]{\dd t}  + {\bf r}^{(1)}(t){\dd t} \, , \\
	\text{Level 3:} &\dd {\bf r}^{(1)}& = {\bf M^{(2)}}[{\bf y}; {\bf r}; {\bf r}^{(1)}]{\dd t}   + {\bf r}^{(2)}(t){\dd t}  \, , \\
    &\hspace{1cm}& \ldots \ldots \nonumber\\
	\text{Level L:} & \dd {\bf r}^{(L-2)} & =  {\bf M^{(L-1)}}[{\bf y}; {\bf r}; {\bf r}^{(1)}; \ldots; {\bf r}^{(L-2)}]{\dd t} + {\bf{Q}\dd\bf{W}} \, ,
\end{eqnarray}
\label{eq:MLLR}
\end{subequations}
\end{linenomath*}
where, as before, $d{\bf W}$  is an independent Gaussian white noise process; and ${\bf Q}$ is the lower triangular matrix obtained by the Cholesky decomposition of the zero-lag correlation between the last-level residuals (same as in Sec. \ref{subsec:AddLR}).
In our results, the residuals at the second level ${\bf r}^{(1)}(t)$ become sufficiently decorrelated in time (according to the stopping criterion) and are well approximated by the spatially correlated white noise, so we used $L = 2$. 

Note that for the prediction experiments in Sec.~\ref{sec:results}, the initial conditions for the residuals at various levels need to be determined in strictly ``no-look-ahead'' procedure, i.e., only using the  model coefficients and the time history of ${\bf y}$ prior to the forecast start time.
E.g., we would need to begin from time instant $(t-1)$ to make forecasts from time instant $t$, when using a model with one extra level (see Appendix B in \citeA{MSM2015}) and initializing ${\bf r}(t-1)=({\bf y}(t)-{\bf y}(t-1))/\dd t-{\bf Ly}(t)$.
To obtain numerical results we have used the publicly available Stochastic Modeling and Prediction Toolbox (see Acknowledgments).  

\subsection{Artificial neural network (ANN)}
\label{subsec:ANN}
A neural network is a computational architecture loosely based on the biological networks of neurons in the human brain.
Each neuron is an instance of an activation function (e.g., linear, binary, hyperbolic, sigmoid) that operates on the weighted sum of its inputs with some bias added into it.   
Multiple neurons can be combined into distinct neural network architectures. 
A feed-forward Artificial Neural Network (ANN) is among the most common \cite{nielsen2015neural}. 
It is composed of multiple layers of neurons so that outputs from one layer are the inputs to the next layer, with the ultimate goal to approximate the right functional relationship between the inputs and outputs. 
In a dense network, each neuron in a layer receives inputs from all the neurons in the previous layer and, thus, exhibits a compact set of connections between the available neurons, inputs, and outputs. 
Each connection in the network is characterized by its weight, and the goal of an ANN is to optimize them using a suitable loss function and optimization algorithm.
An ANN is defined by a few hyper-parameters controlling its performance (e.g., number of hidden layers, number of neurons per layer, activation function, optimizing function, optimizer). 
To adjust the hyper-parameters to the optimal values is a non-trivial task (due to the size of hyper-parameter space), and the adjustment is mostly based on trial-and-error testing that adjusts the model complexity to the amount of data available and the complexity of the problem.

We implemented the ANN using Tensorflow, from Keras Google API, which takes the state of the PCs at time $t$ as the input and returns the state at time $(t+1)$ as the output.
For our training data, the best performing ANN contains two hidden layers of neurons, each with $100$ neurons, hyperbolic tangent as the activation function, \textit{Adam} -- a first-order gradient-based optimizer for stochastic objective functions -- as the optimizer, and mean absolute error as the loss function.
We tried several combinations of the hyper-parameters -- hidden layers, activation function, and optimizer -- and picked up their optimal combination based on tracking the loss function and the naked-eye perception of the model results.
Here we tested ANNs with up to $4$ hidden layers, each with 100, 150, or 200 neurons; linear, elu, relu, sigmoid, and tanh as the activation functions; and RMSprop, SGD, Adagrad, and Adam as the optimizers.
Our intermediate complexity of the model and the training over a prolonged dataset (i.e., $40K$ records) help us avoid over-fitting.
As a sanity check, we trained ANN as $dy/dt = ANN(y(t))$ with the `linear' activation function and compared the outcomes with LR; we found the results to be very similar to each other.

To improve the ANN forecasts further, we added a spatially correlated white noise to the ANN forecasts to account for the residuals, similar to LR in (\ref{eq:SLR_with_white_noise}).
This model is abbreviated as ANN-AWN and is given as:
\begin{linenomath*}
\begin{equation}
    \mathbf{y}(t+1) = ANN(\mathbf{y}(t)) +  \bm\zeta; \, ,
    \label{eq:ANN+Noise}
\end{equation}
\end{linenomath*}
where $\bm\zeta \sim \mathcal{N}(\mathbf{0},\mathbf{Q^TQ})$, and $\mathbf{Q}$ is the lower triangular matrix obtained by Cholesky decomposition of the covariance of ANN residuals $\mathbf{r}(t) = \mathbf{y}(t+1) - ANN (\mathbf{y}(t))$.
Note that it is also possible to train ANN to predict the perturbation $\dd \mathbf{y}$ for a given state $\mathbf{y}(t)$ -- as done in \citeA{chattopadhyay2020dataNP,gmd-11-3999-2018} -- but this approach resulted in unstable solutions for long integrations.

\subsection{Long Short Term Memory (LSTM) Model}
\label{subsec:LSTM}
LSTM models belong to the class of recursive NNs and function by passing information from previous timesteps to calculate the next timestep when used iteratively.
Because these models hold essential dynamical information between the successive time steps, they account for long-time correlations between the model states.
This is a significant advantage over ANNs, for an application with significantly autocorrelated time series.
Like ANNs, LSTMs can also be upgraded using spatially-correlated white noise (abbreviated as LSTM-AWN) with the noise parameters inferred using the LSTM residuals. 

We used Keras Google API to implement a two-hidden-layered densely-connected LSTM configuration with $100$ neurons in each layer.
The model was trained using mean absolute error as the loss function, ``Adam'' as the optimizer, hyperbolic tangent as the activation function, and the whole $400$K days as the training length.
Unlike ANNs, the LSTM model takes the state of the PCs at five previous time steps as the input and produces the next state as the output -- the so-called `look back' hyperparameter is $5$.
Higher values of look back did not improve model performance significantly, but the overall computational cost of training/prediction increases many folds (note that LSTM is optimized for taking into account long-time correlation effects by construction).
For mini-batches, we used $32$, $644$, and $128$ as its potential values and found $32$ to be optimal -- amounting to nearly one year of observations.
For all other hyperparameters, we used the same hyperparameter search space as described in ANN, and the final values were chosen after testing their different combinations and analyzing the resulting model performance on the training data.
\subsection{Hybrid Modeling}
\label{subsec:Hybrid}
The hybrid model that combines LR, which conveys the linear dynamics, with the deep-learning models -- used as a non-linear correction, state dependence, and memory term -- may be more skilled than the standalone implementation of these methods.  
Such a hybrid model can potentially also preserve some core dynamics of the system and may also benefit from simpler algorithms and architectures in the spirit of theory-guided data science \cite{karpatne2017physics}.

We trained the deep-learning models (say, $f$) from the previous sections (ANN and LSTM) to emulate the LR residuals $\mathbf{r}(t) = \dd \mathbf{y}(t)/\dd t - \mathbf{Ly}(t)$ and, thus, augment the LR output as:
\begin{eqnarray}
\label{HybridNN}
\dd\mathbf{y}(t)  & = & \mathbf{L}\mathbf{y}(t){\dd t} +\mathbf{r}(t){\dd t}, \nonumber
\\
\mathbf{r}(t)& = &f(\mathbf{r}(t-1),\mathbf{y}(t-1); \mathbf{r}(t-2),\mathbf{y}(t-2); \ldots; \mathbf{r}(t-l),\mathbf{y}(t-l)) +\bm{\xi},
\end{eqnarray}
where $l$ is the `look back' hyperparameter for LSTM; for ANN, it is equal to $1$ by construction.
For LSTM, we set $l=3$ after checking the LSTM-hybrid model performance on the training data for $l = 1, 2, \dots,5$ and finding that the model performance does not improve beyond $l=3$.
The model learning proceeds in three successive steps as follows: (1) LR is used to estimate $\mathbf{L}$, (2) the resulting residual $\mathbf{r}(t)={d\mathbf{y}}/dt -\mathbf{L} \mathbf{y}(t)$ is modeled by $f(\mathbf{r}(t-1),\mathbf{y}(t-1); \ldots; \mathbf{r}(t-l),\mathbf{y}(t-l))$ that accounts for non-linear correction, state dependence and possibly memory effects, (3) the final residual from deep-learning minimization is approximated by a spatially correlated white noise process $\bm \xi \sim \mathcal{N}(0,\mathbf{Q}^{T}\mathbf{Q})$.
This procedure can be interpreted as incorporating state-dependent noise $\mathbf{r}(t)$, and, as said previously, it is implemented for the two deep-learning methods (described in the previous sections), referred to as LR-ANN-AWN and LR-LSTM-AWN. 
We have also evaluated versions of Eq.(\ref{HybridNN}) with no stochastic forcing, i.e., without the white noise term, that are abbreviated as LR-ANN and LR-LSTM. 

Note that unlike their standalone implementations, the deep-learning models here take both the state $\mathbf{y}(t)$ and residual $\mathbf{r}(t)$ as inputs, and return the residual $\mathbf{r}(t+1)$ as the output.
We tested this configuration against several others and found that the current setup performs better than the others on both training and test datasets. 
\section{Results}
\label{sec:results}

In this section, we consider each model assessment metric separately, summarize them, and report the outcome for all models.
For the majority of the assessment metrics, we use the reconstructed spatio-temporal streamfunction field ($\psi^{fcast}_1$), obtained by multiplying the forecasted PCs ($y_i$'s; i = 1,2,3,\ldots,150) with the respective EOFs ($\phi_i$). 
We obtain PC forecasts corresponding to a set of initial conditions.
The exact number of initial conditions and the lengths of the forecasts differ for the assessment metrics and are provided in the detailed descriptions.
Additionally, for stochastic methods, we obtain an ensemble of $100$ realizations for each initial condition and calculate the ensemble mean.
The reference truth ($\psi^{ref}_1$) used for assessing the model outputs belongs to the reduced space spanned by the $150$ EOFs/PCs.
Below, we present the assessment metrics and detailed analyses of the models. \subsection{Root Mean Square Error (RMSE)}
\begin{figure}[!ht]
    \centering
    \includegraphics[width=\linewidth]{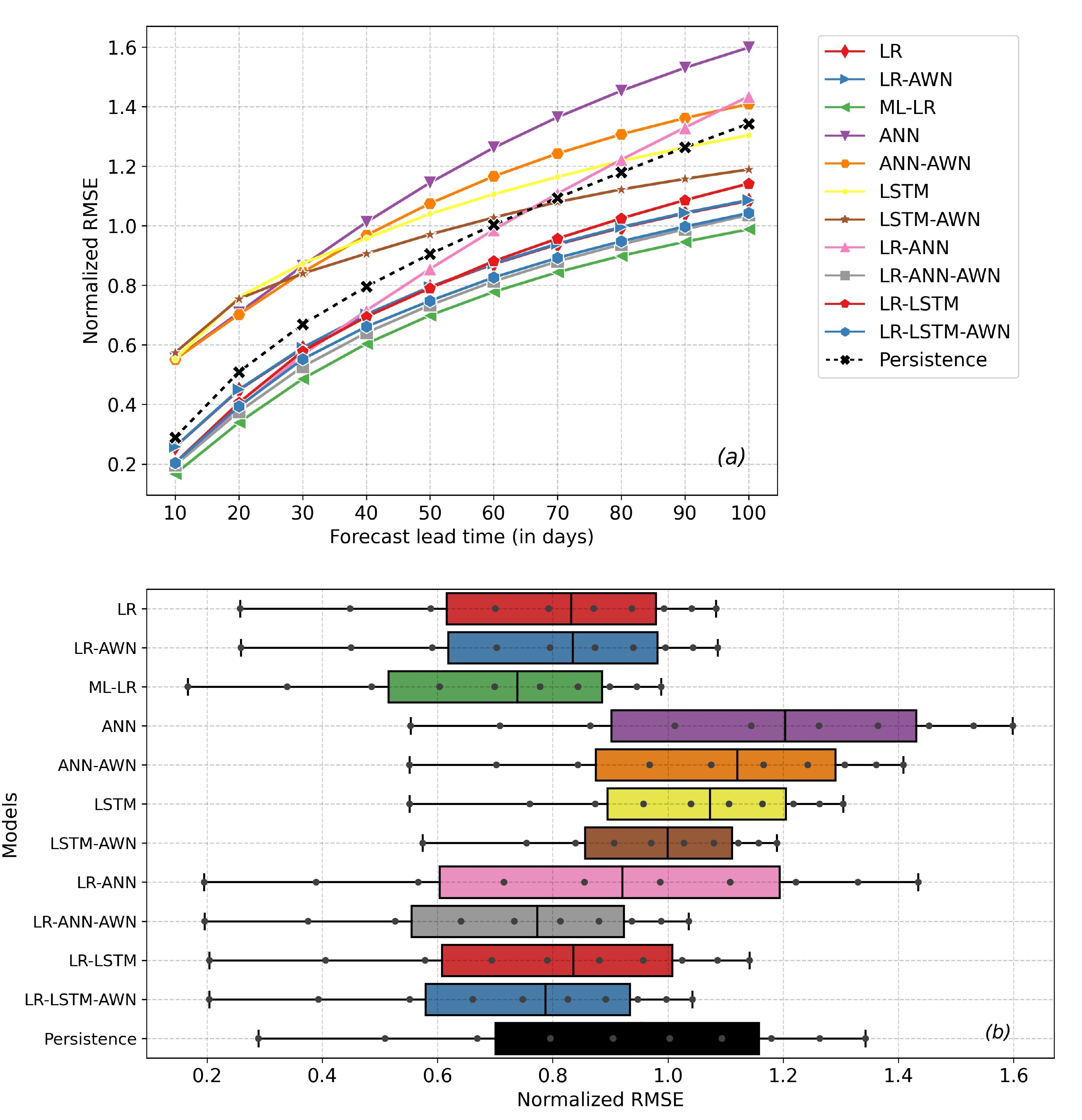}
    \caption{
    Comparison of the RMSE curves for all the models used in this work: (a) line plot, and (b) box plot. 
    The RMSEs are normalized by the standard deviation of the test dataset. 
    The line plot compares the RMSE for all models at each forecast lead time, whereas the box plot provides a visualization of the statistical range and min-max values of the RMSE over $100$-days forecast.  
    }
    \label{fig:RMSE}
\end{figure}

The RMSE is given as
\begin{linenomath*}
    \begin{equation}
        E(t) = \sqrt{\frac{1}{m^2}\sum_{i=1}^m\sum_{j=1}^m \Big(\psi^{ref}_1(x_i,y_j,t) - \psi_1^{fcast}(x_i,y_j,t)\Big)^2} \, ,
    \end{equation}
\end{linenomath*}
and its time series describes the basin-averaged mismatch between the reference and emulated streamfunction snapshots.
For perfect forecasts, the RMSE should be zero.
However, in practice, RMSE is small for short forecast lead times and grows 
until it becomes saturated, as the forecast and reference truth become uncorrelated at long forecast lead times.
We considered the state of the PCs at each of the $10K$ records of the test dataset ($10K$ records correspond to $100K$ days) as an initial condition and obtained $10$ records long (i.e., $100$ days) forecasts for all of them; therefore, we used a total of $9990$ initial conditions (as we would not have the reference data for the last $10$ initial conditions).
Next, we computed RMSEs for each of these forecasts followed by the mean RMSE over all initial conditions; this provides a $10$ records long time series for each model (Fig. \ref{fig:RMSE}a).
The motivation for using this forecast length is to study the error growth for short-term forecasts (e.g., on seasonal time scales) as opposed to a change of the mean fields using long-term simulations (e.g., decadal-to-centennial). 
The short-term predictability for a single initial condition can depend strongly on the underlying flow regime, such as defined by the western boundary current position. 
It is therefore important to take the average RMSE over a number of initial conditions. 

For comparison purposes, we also considered the ``persistence'' model, where the memory effect is the strongest, and the model state remains constant -- equal to the initial condition.  
The persistence RMSE time series, therefore, characterizes a ``constant state'' with the absence of a dynamic model.

For $100$ days, ML-LR exhibits the best performance with the lowest mean RMSE, followed by the stochastic augmentations of the hybrid models, i.e., LR-ANN-AWN and LR-LSTM-AWN (Fig. \ref{fig:RMSE}a). 
The deterministic hybrid model RMSEs are similar or worse than LR and its white noise extension; LR-ANN is worse compared to LR-LSTM as the RMSE for the former grows more steeply with the lead time and gets even higher than the Persistence for a lead time beyond $70$ days. 
This suggests that the additive noise is improving the deterministic hybrid methods (clearer from Fig. \ref{fig:RMSE}b), as the residuals in such models are less correlated and closer to white noise.
We also observe that the standalone ANN, both in its deterministic and stochastic version, performs the worst, with its RMSEs being higher than the Persistence at all lead times.
The standalone LSTM also performs similarly poorly (yet better than its ANN counterpart), but its stochastic version produces RMSEs lower than the Persistence at high lead times. 
Nonetheless, a comparison of standalone and hybrid implementations of deep learning methods encourages us to use ANN/LSTM as a nonlinear corrector term rather than using them to represent the complete dynamics.

On the other hand, LR and LR-AWN belong to the middle of the RMSE spectrum and show similar performances (see the box plot, Fig. \ref{fig:RMSE}b).
The similar performances of these models is due to the inability of the noise component to account for the coupled dynamics contained in the LR residuals, which are modeled more efficiently using an extra regression level (as in ML-LR) or deep-learning methods (as in the hybrid methods).  

Overall, we conclude that (i) ML-LR and the stochastically augmented hybrid models show better performance than the standalone implementation of LR and deep-learning models, probably, because the former types include all three major components of a reliable model: core dynamics, memory effects, and model errors accounted by stochastic noise; (ii) adding simple additive noise to the hybrid models significantly improves their performance. \subsection{Anomaly Correlation Coefficient (ACC)}
Next, we diagnose the correlation between the forecasted ($100$-days-long forecasts) and the reference truth spatio-temporal streamfunction datasets on spatial and temporal domains, referred to as ASCC and ATCC, respectively.
We call this anomaly correlation because we deal with mean-subtracted PCs, and, therefore, the resulting physical fields possess zero mean. 

\begin{figure}
    \centering
    \includegraphics[width=\linewidth]{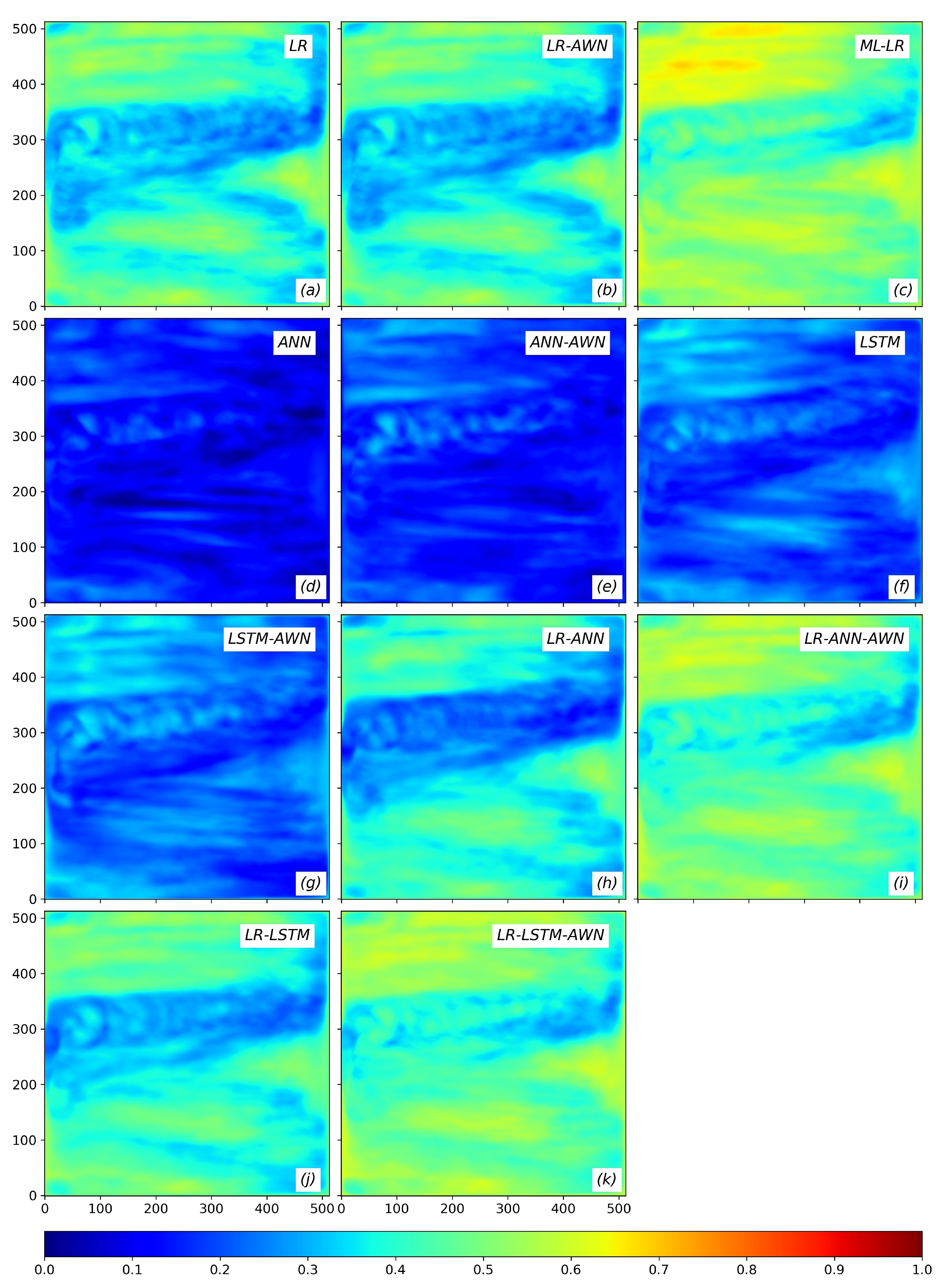}
    \caption{
    ATCC for $100$-day forecasts obtained using different models.
    The color scale presents the correlation values.
    }
    \label{fig:ATCC}
\end{figure}

ATCC is a grid-point-wise zero-lag cross-correlation between the forecast and the reference truth over all lead times. 
It, therefore, gives us a gridded map with the zero-lag cross-correlation value between the forecast and the reference truth for each grid location.
Like RMSE, the ATCC map is computed for each of the $9990$ initial conditions followed by their average (Fig. \ref{fig:ATCC}).

\begin{figure}
    \centering
    \includegraphics[width=\linewidth]{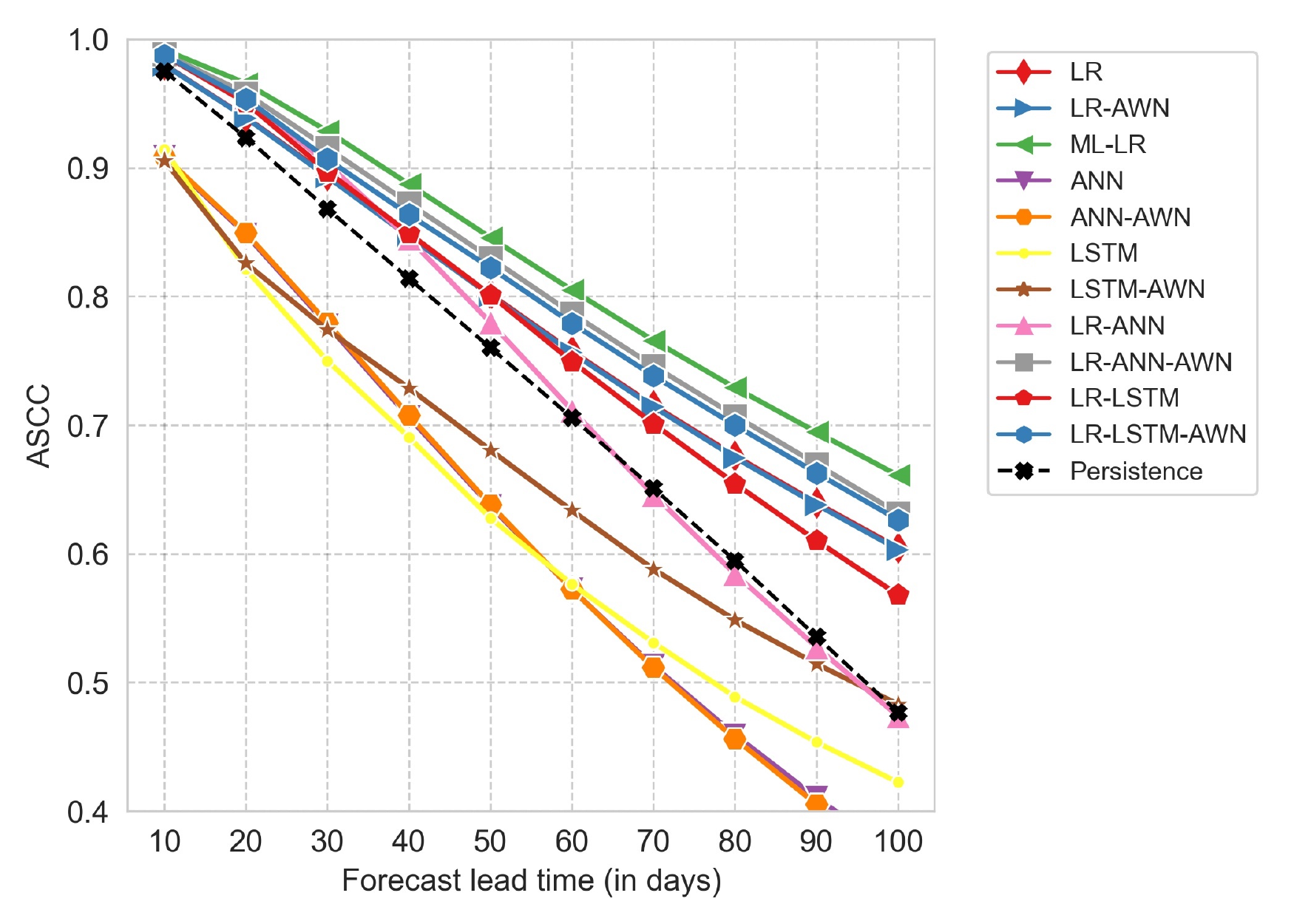}
    \caption{Comparison of ASCC for all models.}
    \label{fig:ASCC}
\end{figure}

ASCC is the cross-correlation between the spatial snapshots of reference truth and forecast at each lead time, say, $t$.
The snapshots of forecast and the reference truth are reshaped to a vector before cross-correlation.
ASCC therefore returns a time series of length $100$ days (the maximum lead time) for each initial condition, and we report their average (Fig. \ref{fig:ASCC}).  

\begin{linenomath*}
    \begin{subequations}
        \begin{eqnarray}
            &ATCC(x,y)&= \Bigg\langle\Bigg(\frac{\psi_1^{fcast}(x,y,t) - \langle \psi_1^{fcast} \rangle(x,y)}{\sigma(\psi_1^{fcast})(x,y)}\Bigg)\cdot \Bigg(\frac{\psi_1^{ref}(x,y,t) - \langle \psi_1^{ref} \rangle(x,y)}{\sigma(\psi_1^{ref})(x,y)}\Bigg)\Bigg \rangle ,\\
            &ASCC(t)&= \overline{\Bigg(\frac{\psi_1^{fcast}(x,y,t) - \overline{\psi_1}^{fcast}(t)} {\sigma(\psi_1^{fcast})(t)}\Bigg)\cdot \Bigg(\frac{\psi_1^{ref}(x,y,t) - \overline{\psi_1}^{ref}(t)}{\sigma(\psi_1^{ref})(t)}\Bigg)},
        \end{eqnarray}
        \label{eq:ACC}
    \end{subequations}
\end{linenomath*}
where $t$ is the forecast lead time, $\overline{(.)}$ and $\langle.\rangle$ indicate the spatial and temporal averages, respectively, and $\sigma$ refers to the standard deviation.
Essentially, ATCC conveys temporal similarity between the forecasted and reference truth over all lead times, as it computes the grid-point-wise temporal correlation between the time series of the two datasets averaged for all initial conditions. 
In contrast, ASCC exhibits the spatial structure similarity between the two datasets, as it computes the snapshot-wise correlation between the two fields at a given lead time, normalized and then averaged for all initial conditions. 

The ATCC maps show that the ML-LR forecasts are the best, followed by the stochastic hybrid models (Fig. \ref{fig:ATCC}c,i,k).
For most of the models, the correlations are generally higher in the gyre regions than in the eastward jet region, which is justified since the latter area is more turbulent.
However, the stochastically-improved hybrid methods and ML-LR provide higher correlation in the jet region than those of the non-stochastic hybrid models (Fig. \ref{fig:ATCC}h,j).
The pure deep-learning methods and their stochastic extensions (Fig. \ref{fig:ATCC}d-g) fail to reproduce the variabilities entirely, thus, resulting in significant basin-wide dissimilarity with the reference dataset.
The ATCC maps of LR and its white noise extension (Fig. \ref{fig:ATCC}a-b) show similar basin-wide correlations, and these are similar to the deterministic hybrid models (Fig. \ref{fig:ATCC}h,j). 
However, small correlations along the jet region are more pronounced in LR-AWN. 

A comparison of the ASCC time series (Fig. \ref{fig:ASCC}) is telling a similar story.
The pure deep-learning methods and their stochastic augmentations show the worst structural similarity with the reference truth (lower than the Persistence at all lead times), whereas ML-LR provides the highest correlation, followed by the noise-augmented hybrid models.
An inspection of the correlation decay rates of the models involving LR suggests that, on average, all of them possess a similar decorrelation rate, but the deterministic hybrid models decay faster than the others on longer lead times with LR-ANN decaying similar to the Persistence baseline.
This suggests that introducing noise in the hybrid models improves both temporal and spatial characteristics.

Overall, we conclude that, on short forecast time scales, both ML-LR and noise-augmented hybrid methods are most realistic regarding the spatial and temporal evolution. 
The pure deep-learning models and their stochastic extensions perform poorly. \subsection{Climatology and Variance}
\begin{figure}
    \centering
    \includegraphics[width=0.95\linewidth]{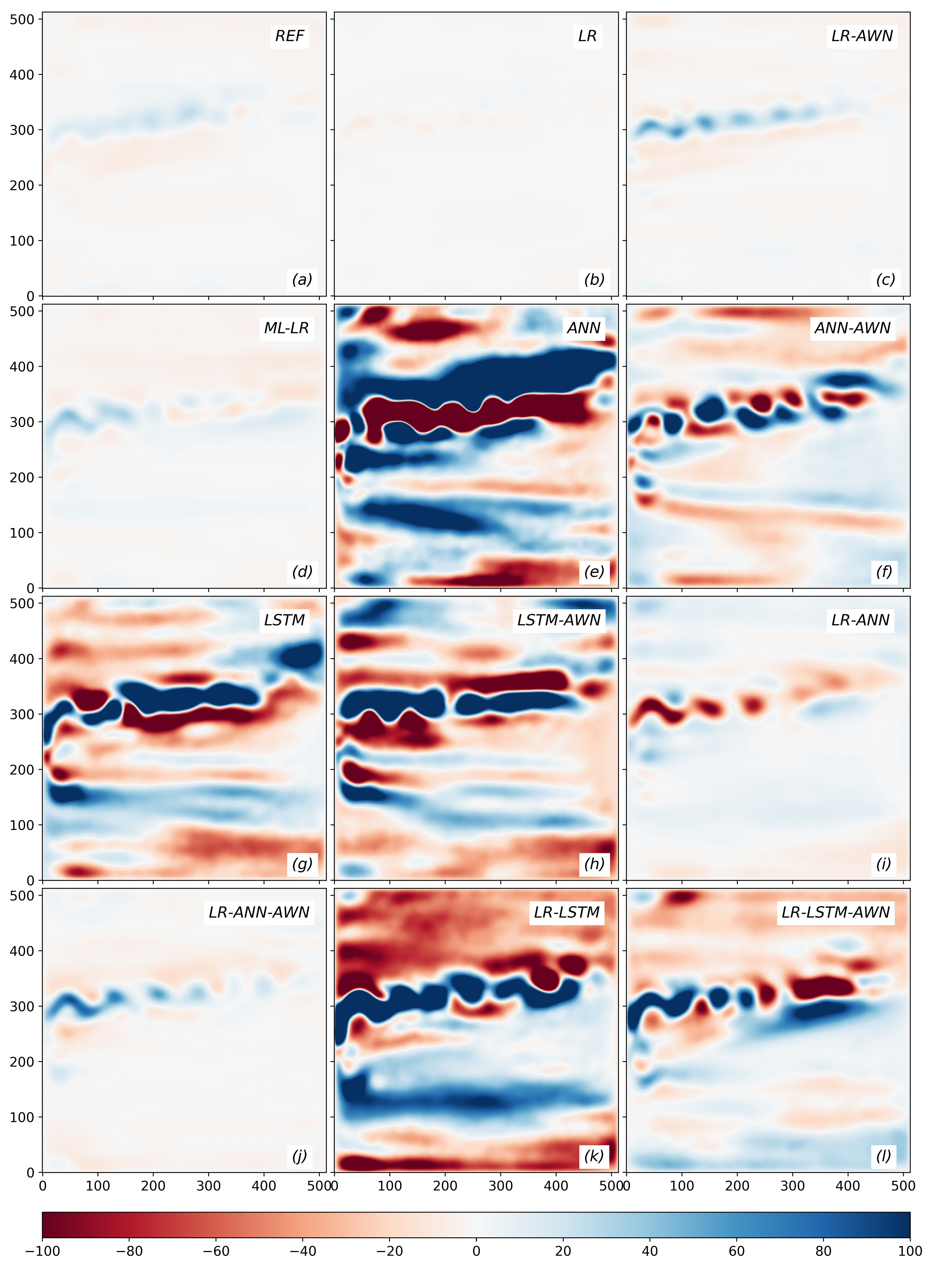}
    \caption{
    The climatology for $200$K days randomly chosen data sample from the $500K$ days reference dataset (REF; top left panel); and the $200K$ days forecasts produced by different models.
    The  range of the climatological bias $[-100, 100]$ should be compared to the range $[-10^3, 10^3]$ of the time-mean state of streamfunction, see Fig. \ref{fig:flow_snapshot}b for a streamfunction snapshot; non-dimensional units.}
    \label{fig:climatology}
\end{figure}

\begin{figure}
    \centering
    \includegraphics[width=0.95\linewidth]{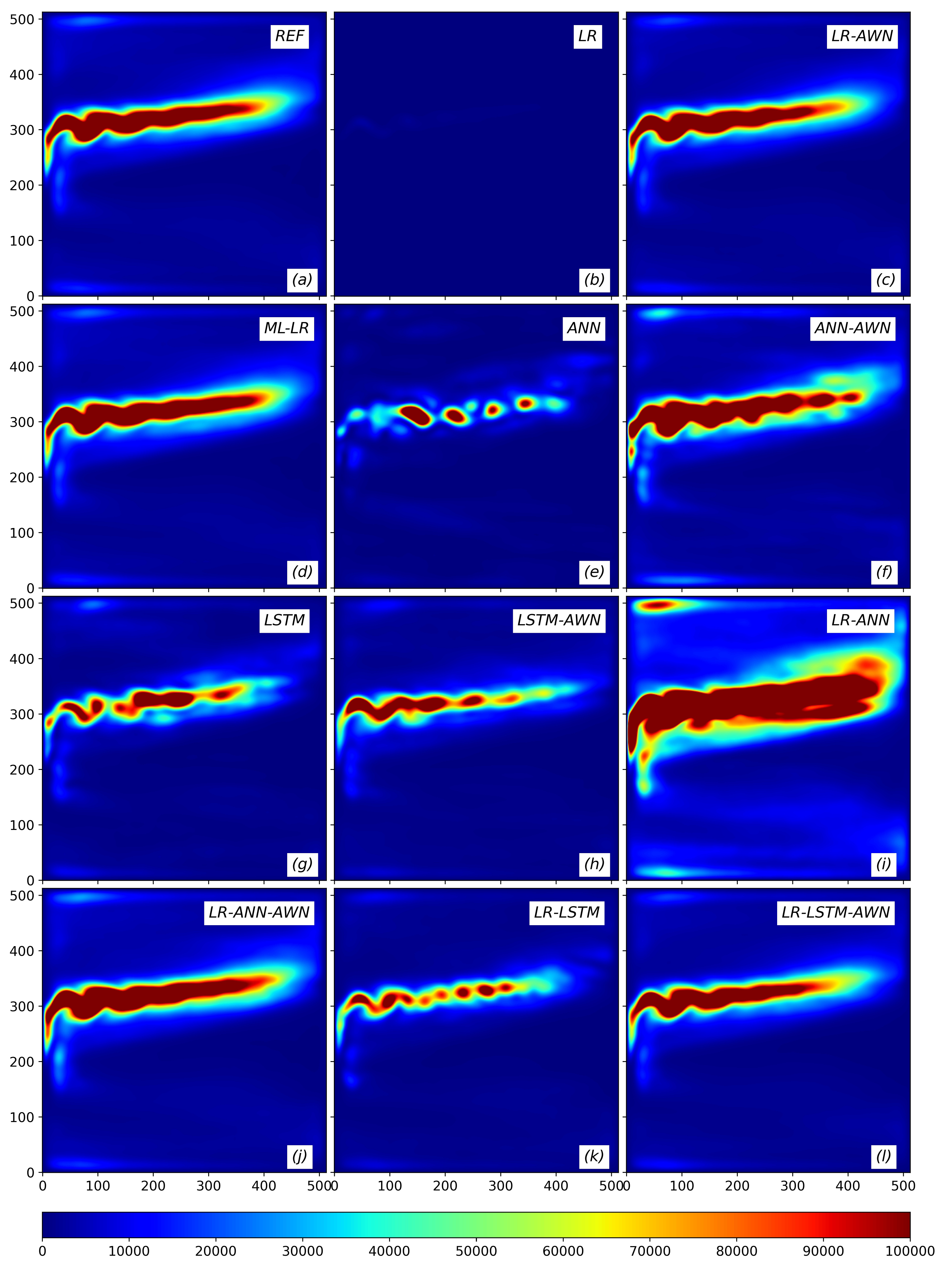}
    \caption{
    The temporal variance for $200$K days randomly chosen data sample from the reference data (REF; top left panel) and $200K$ days forecasts produced by different models.
    Like climatology, the variance is also in non-dimensional units.
    }
    \label{fig:variance}
\end{figure}

Here, we diagnose the mean and variance of the forecasted streamfunction field along the temporal domain: 
\begin{linenomath*}
    \begin{subequations}
        \begin{eqnarray}
            \Big\langle \psi_1^{fcast} \Big\rangle &=& \frac{1}{N}\sum_{n=1}^N \psi_1^{fcast}(x,y,t_n) \, ,\\
        \sigma^2 (\psi_1^{fcast}) &=& \frac{1}{N}\sum_{n=1}^N \Big(\psi_1^{fcast}(x,y,t_n) - \Big\langle \psi_1^{fcast} \Big\rangle\Big)^2 \, .
        \end{eqnarray}
    \end{subequations}
\end{linenomath*}
The mean field $\langle \psi_1^{fcast} \rangle$ is also referred to as the ``climatology''.
However, unlike the operational forecasts, the reference climatology is zero as we have performed an SVD of the mean-subtracted streamfunction field. 
Therefore, the reconstructed streamfunction $\psi_1^{fcast}$ should show a climatology of zero.
We used these diagnostic metrics to characterize long-timescale forecasts, over $20K$ records, i.e., $200K$ days or nearly $547$ years.
As results are independent from the initial conditions, we perform simulations from a single initial condition and use only one stochastic realization, wherever applicable.
Due to a much longer record of the reference streamfunction compared to the forecast length, we expect a small nonzero value of the predicted time-mean streamfunction field (see Fig. \ref{fig:climatology}a for reference dataset of the same length), and its value can serve as a reference for the temporal bias introduced by different models.
On the other hand, the variance map would characterize the extent of jet reproduction by different methods, as it is the most turbulent and possesses maximum fluctuations in the entire domain (Fig. \ref{fig:variance}a).

The climatology maps suggest that the LR shows the least temporal bias among all methods (Fig. \ref{fig:climatology}b).
However, this is because the LR output decays to zero after a short lead time, and the PCs exhibit near-zero mean despite the poor forecasts.
The multi-level formalism (Fig. \ref{fig:climatology}d) shows the second smallest bias, with stable and non-zero forecasts at all lead times.
The AWN extension of LR produces the next overall small bias among the LR and its stochastic extensions (Fig. \ref{fig:climatology}c). 

All standalone deep-learning methods (Fig. \ref{fig:climatology}e-h) produce a relatively large bias in the forecasted streamfunction field.
For ANN, the deterministic version produces a higher bias than the stochastic one, whereas LSTM produces a high bias irrespective of the noise.
The large bias is primarily because these methods produce a significant drift in the modeled PCs, and this results in a large non-zero temporal mean which also reflects in the reconstructed streamfunction field after multiplying with the EOFs.
Among the hybrid models (Fig. \ref{fig:climatology}i-l), the ANN hybrid models generate a smaller basin-wide bias than the LSTM hybrids, and the stochastically improved hybrid models produce a smaller bias than their deterministic counterparts.
The latter suggests that the induced drift in the modeled PCs can be contained to some extent by adding stochasticity, which nudges the PCs back towards the reference truth trajectory.

Analyzing the spatial map of grid-point-wise temporal variance of the model outputs (Fig. \ref{fig:variance}), we found that LR-AWN, ML-LR, and the stochastically augmented hybrid models (Fig. \ref{fig:variance}c,d,j,l) best reproduce the reference jet variability (Fig. \ref{fig:variance}a).
The non-stochastic ANN-hybrid model (Fig. \ref{fig:variance}i) produces correct but extra stretched jet variability in the north-south direction and around the eastward extension, whereas the deterministic LSTM-hybrid model (Fig. \ref{fig:variance}k) produces irregular and overly narrow region of jet variability. 
Among the standalone deep-learning models (Fig. \ref{fig:variance}e-h), the deterministic versions of both ANN and LSTM produce insufficient jet variabilities, with ANN being worse than LSTM. 
But, when augmented with white noise, both produce more variabilities along the jet region and are nearer to the reference truth, especially ANN-AWN.
The LR fails to produce any variability in the basin (Fig. \ref{fig:variance}b) as its outputs decay to zero after a short forecast lead time.

Combining the climatology and variance results, we conclude that (i) on long time scales, LR-AWN, ML-LR, and the stochastically augmented hybrid methods best reproduce the jet variabilities with a small drift in the resulting flow field (approx. $2-3\%$ of the original mean field);
(ii) the standalone deep-learning implementations infuse a relatively higher bias in the climatology and are inefficient at reproducing the turbulent jet characteristics on long time scales;
(iii) the LR model cannot produce climate-like forecasts due to its dissipative nature, and therefore all other models fare better than it on long timescales.
 \subsection{Frequency map}
\begin{figure}
    \centering
    \includegraphics[width=0.95\linewidth]{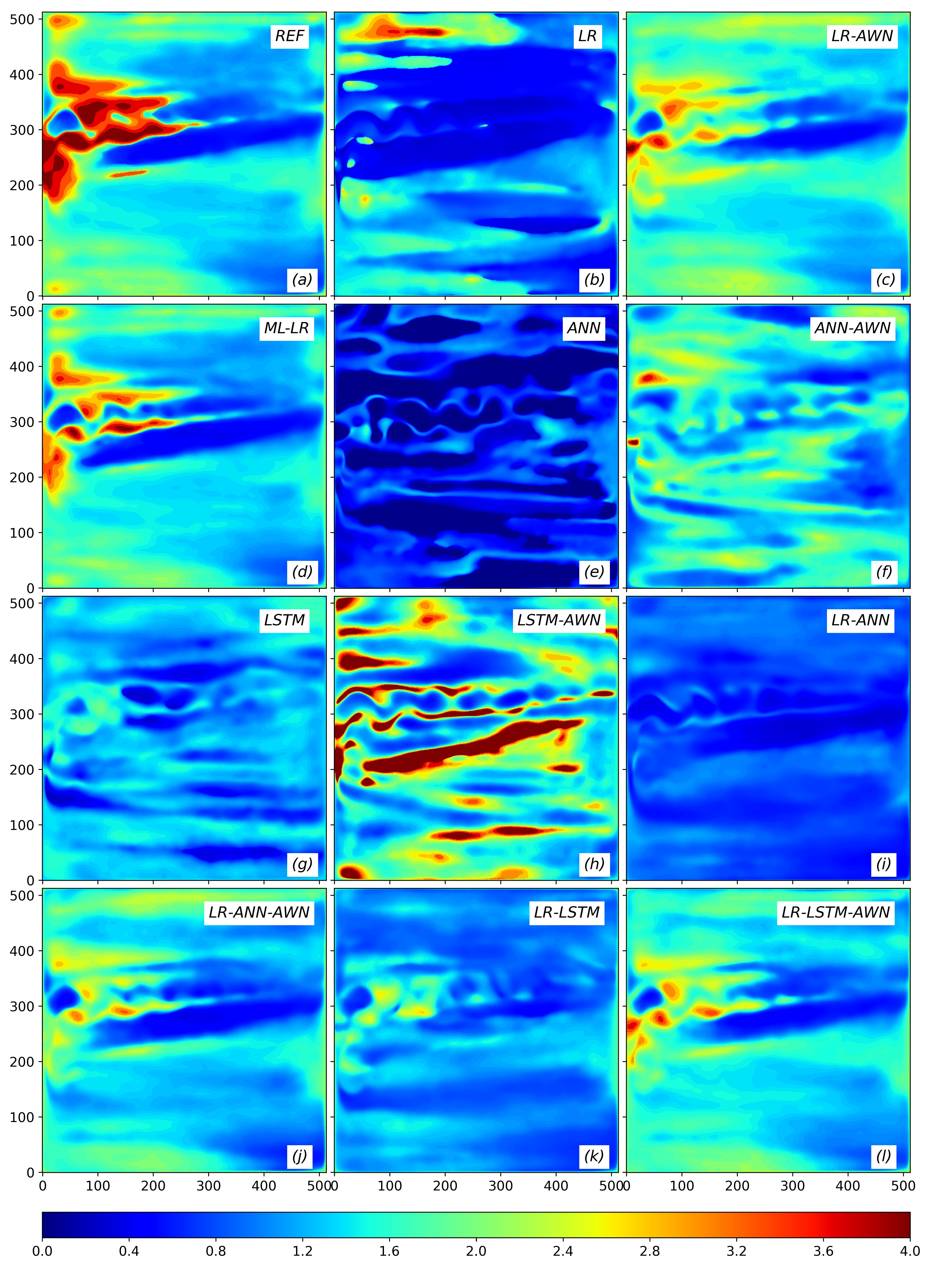}
    \caption{
    Frequency maps for $200$K days long randomly chosen data sample from the reference data (REF; top left panel) and for the $200K$ days forecasts produced by different models.
    The units of frequency are in cycles/year.
    }
    \label{fig:frequency}
\end{figure}

Here, we consider \textit{frequency maps} of the emulated long-timescale solutions (same as the one used in the previous section) to quantify their spectral frequency characteristics.
For each model output, the frequency map is obtained by diagnosing the frequency value locally (i.e., for each grid cell), as given by the inverse of the decorrelation time scale of the forecasted streamfunction field.
The decorrelation time scale is determined as the lag at which the autocorrelation drops by a factor of $e$ from the zero-lag value.
Because we repeat this calculation for each spatial location, we get a gridded frequency map of the size $m \times m$.
Overall, we expect higher-frequency variability along the eastward jet and in boundary regions, due to the vigorous eddy activities, and low-frequency variabilities elsewhere.
This is demonstrated in the reference frequency map (Fig. \ref{fig:frequency}a) obtained for a randomly chosen $200K$ days long data sample from the reference dataset.

The results suggest that ML-LR most closely resembles the reference frequency map followed by LR-AWN and the two stochastic hybrid models (Fig. \ref{fig:frequency}d,c,j,l) . 
While ML-LR produces the correct frequency patterns in the gyre regions, the magnitude is lower in the jet region when compared to the reference truth.
LR-AWN and the stochastic hybrid models also reproduce the frequencies in the gyre regions but are less accurate in the jet region, with the frequency magnitude even lower than ML-LR in this region.

On the other hand, the deterministic hybrid models (Fig. \ref{fig:frequency}i,k) fail to describe the characteristic frequencies throughout the domain and produce frequency maps mostly dominated by low frequencies -- more so for the deterministic ANN hybrid model. 
The same is also true for the standalone ANN implementations (Fig. \ref{fig:frequency}e,f) with the stochastic one being better as they show higher frequencies in the domain but not following the correct pattern.
Such low-frequency dominated maps suggest that the individual PC outputs lack correct high-frequency contributions. 
The LSTM-only models also produce incorrect frequency maps irrespective of the noise (Fig. \ref{fig:frequency}g,h). 
The deterministic LSTM variant exhibits patterns of low and intermediate frequencies in the entire basin, whereas the stochastic variant produces patches of high frequencies both in the jet and gyre regions and do not resemble their reference truth.
This is because the deterministic LSTM outputs (for all $150$ PCs) are dominated by low frequencies, and adding noise to them produces somewhat large-amplitude (therefore, higher variance) but much high-frequency-dominated outputs for most of the PCs, which ultimately leads to high frequencies in both the jet and the gyres regions when multiplied by the EOFs.

LR completely misses the high-frequency variability around the jet (Fig. \ref{fig:frequency}b), as the solution decays to zero, although it still manages to reproduce the low-frequency variations in the two gyres to a certain extent. 

We conclude that (i) as seen for the climatology and variance, LR-AWN, ML-LR, and the stochastic hybrid models perform best regarding frequency characteristics of long-timescale solutions; (ii) the deterministic hybrid models fail to correctly reproduce the frequency content despite their low climatological bias and nearly correct variance pattern; (iii) standalone deep-learning methods produce the most inaccurate and physically unjustified frequency maps, especially ANN. \subsection{Forecast Horizon}
Here, for each model, we estimate the \textit{forecast horizon} which is the time scale for which the model produces stable and non-zero forecasts.
This information is vital for deciding on the applicability of a method in short-/long-term forecasts.  
When the system matrix is available, as in the LR, the forecast horizon is given by the inverse of its maximum eigenvalue. 
For the other methods, it is computed using the model outputs, which, for long-timescale forecasts, either saturate to a steady-state value or provide non-zero and stable solutions up to the maximum lead time (equal to $20K$ records, i.e., $200K$ days or $547$ years).
In the first case, the forecast horizon for each PC is given by the time beyond which the solution is trapped in a small-amplitude range, i.e., a small threshold value.
The overall forecast horizon is, then, set equal to the minimum of the individual PC horizons.
In the second case, the model forecast horizon is set to $\infty$ because such models can produce stable non-zero forecasts on any finite lead time. 

Only LR belongs to the first category, whereas all the other methods belong to the second (Table \ref{tab:FH-RTC}, column 2). 
LR is a dissipative deterministic model, and its outputs decay to zero around its forecast horizon.
Therefore, this model is not suitable for climate-type applications.
It is also worth noting that all stochastically-forced models possess infinite forecast horizon.
This implies that using random variables alongside the PCs ensures the injection of energy (in the form of small-scale variabilities) that prevents decay of the solutions and, at the same time, keeps the solution bounded and stable.

\begin{linenomath*}
\begin{table}[!ht]
    \centering
    \begin{tabular}{p{3cm}|p{3cm}|p{3cm}|p{3cm}}
         Method & Forecast Horizon (in years) & Training Time (in seconds) & Prediction Time (in seconds)  \\ \hline
         LR & $1.3$ & $1$ & $10^{-1}$\\
         LR-AWN & $\infty$ & $1$ & $10^{-1}$ \\
         ML-LR & $\infty$ & $10$ & $1$ \\
         ANN & $\infty,\infty$ & $10^3$ & $1$ \\
         LSTM & $\infty, \infty$ & $10^{4}$ & $10^2$ \\
         LR-ANN hybrid & $\infty, \infty$ & $10^3$ & $1$ \\
         LR-LSTM hybrid & $\infty, \infty$ & $10^3$ & $10^2$ \\
    \end{tabular}
    \caption{
    Forecast horizon and running time complexity analyses for all the models.
    The two values of the forecast horizon for ANN, LSTM, and the hybrid models represent these values for their deterministic and stochastic variants, in the same order.
    However, their training and running times are of the same order of magnitude, so we have reported them only once.
    Both training and prediction times are reported for the codes executed on a $24$ core machine with $64$gb memory.
    For deep-learning-based methods, the training time is sensitive to the number of epochs and the batch size.
    Here, we have reported them for $200$ epochs and the default batch size $32$.
    Note, however, that some of the deep-learning-based methods, e.g., ANN, achieved optimal training in a smaller number of epochs. 
    }
    \label{tab:FH-RTC}
\end{table}
\end{linenomath*} \subsection{Training and prediction time complexity}
\label{sec:RTC}
Here, we discuss computational costs and scalability as the number of degrees-of-freedom is increased.
For each model, we have diagnosed the run times for training and inference and refer to them as the ``training time complexity'' and ``prediction time complexity'', respectively.
In practice, the training is done only once, whereas the predictions are made many times, therefore, it makes sense to look at their time complexities separately.
Note that the complexity estimates should take into account different levels of possible optimization of the models.
Therefore, we wrote all model codes in Python (except ML-LR that is in a publicly available Matlab Toolbox), implemented them on the same hardware, and optimized them to reasonably high levels, including vectorization, function calls, and efficient data structures.
The training complexity estimates do not include data processing and variable declaration/initialization, and only correspond to the time taken for training the models.
The prediction time complexity corresponds to the time taken for producing one realization of a $1000$ records long forecast. 
Since we can only provide estimates for the optimal performance of different methods, we only report the orders of magnitude of the elapsed time (Table \ref{tab:FH-RTC}, column $3$ and $4$).

Among the stochastic augmentations of LR, ML-LR is one order of magnitude more expensive to train and forecast compared to LR and its additive white noise extension. 
The higher complexity of ML-LR is simply due to the extra regression layer.
Similar training and prediction times of LR and LR-AWN are due to the same trained core, i.e., LR coefficients; the white noise parameters in LR-AWN are inexpensive to train.

Mathematically, for LR and its white noise stochastic extension, as the number of PCs (say, $n$) increases, we expect the training time complexity to increase as $\mathcal{O}(n^2)$, which is the size of the trained regression matrix, and the prediction time complexity to increase as $\mathcal{O}(n)$.
Due to the added levels (say, $l$) in the multi-level formalism, the corresponding training and prediction times are expected to increase additionally by $\mathcal{O}(n^2l + nl^2 + nl)$, assuming $n>>l$, and $\mathcal{O}(l)$, respectively.

In the standalone deep-learning class, LSTM is an order of magnitude more expensive to train than ANN, but two orders of magnitude costlier to produce forecasts.
The higher computational cost of LSTM is due to its more complex architecture and a large number of past states passed as input -- set by the `look back' hyperparameter.
However, in the hybrid category, both ANN- and LSTM-based methods follow the same order of training time, but the difference in their prediction cost is the same -- i.e., two orders of magnitude.
A modest difference between the training cost of ANN and LSTM in the hybrid design is due to their simpler architectures and a smaller number of past states required by LSTM to predict the next state.
Nevertheless, in both categories, both ANN and LSTM are $1-3$ orders of magnitude cheaper to produce forecasts than to train them.
This makes them suitable for climate-related applications, as the training is done once, but the predictions are obtained numerous times.

We do not compare the time complexity of the deep-learning-based methods against the LR and its stochastic extensions because the time estimates for both ANN and LSTM depends on their hyper-parameter values, mainly \textit{Epochs} and \textit{Batch size} (they were kept the same for all ANN- and LSTN-based models), which are bound to change for different applications. 
\textit{Epochs} determine the number of passes through the entire training dataset needed to optimize the model parameters; \textit{Batch Size} refers to the number of training samples which needs to be parsed before updating the model parameters.
Due to these hyper-parameters, with a few others such as the optimizer, loss function, activation function, etc., and the black-box nature of NNs, it is hard to determine how exactly the running time complexities scale with $n$.

Overall, we conclude that (i) adding simple stochasticity bears a negligible computational cost, but a more complicated red-type noise addition can increase the training and prediction cost by order of magnitude; 
(ii) Among the deep learning methods, LSTM is equal or more costly to train and forecast than ANN; 
(iii) Both LSTM and ANN can benefit from reduced training time in the hybrid framework than in its standalone implementation, as a less complex network design is required for optimal performance. 
\section{Discussion and Conclusion}
\label{sec:conclusions}

We have presented a comprehensive inter-comparison of Linear Regression (LR), its various stochastic extensions, deep-learning models (ANN and LSTM) and their hybrid formulations with additive-noise (see Table \ref{tab:overview} for an overview), to obtain a low-cost, reduced-order model for complex multi-scale spatio-temporal flow of the upper ocean.
LR has the simplest form and, thus, provides a baseline for assessing the performance of the other models.
The obtained results show that linear models augmented by state-dependent noise and memory effects, either through multi-level regression or deep learning, perform the best across our metrics and tasks to emulate and predict very complex, nonlinear and multi-scale ocean flow.
Convolutional Neural Network (CNN) is another proven methodology in the deep-learning class for emulating image-based datasets but is not considered here as it is well-studied elsewhere and is beyond the scope of this paper as it would be most suitable for structured two-dimensional grids rather than for the application in EOF space. 
It would therefore require a different dataset likely with a much larger number of degrees-of-freedom.

\begin{linenomath*}
\begin{table}
    \centering
    \begin{tabular}{|p{3cm}|p{2cm}|p{1cm}|p{1.5cm}|p{1.4cm}|p{1.3cm}|p{1.2cm}|p{1.5cm}|}
        Method & Abbreviation & Section & Input & Output & Cost function & Memory & State-dependent Noise \\ \hline
        Linear Regression & LR & \ref{subsec:LR} & state & tendency & OLS & NA & NA\\ \hline 
        LR + Additive White Noise & LR-AWN & \ref{subsec:AddLR} & state & tendency & OLS & \xmark & \xmark \\ \hline
        Multi-level Linear Regression & ML-LR & \ref{subsec:MLLR} & state, LR residuals & tendency & OLS & \checkmark & \checkmark\\ \hline
        Artificial Neural Network (+ White Noise) & ANN (- AWN) & \ref{subsec:ANN} & state & state & MAE &\xmark & \xmark\\ \hline
        Long Short Term Memory (+ White Noise) & LSTM (- AWN) & \ref{subsec:LSTM} & state & state & MAE & \checkmark & \xmark \\ \hline
        LR + ANN Hybrid (+ White Noise) & LR-ANN (-AWN) & \ref{subsec:Hybrid} & state, LR residuals & tendency & OLS, MAE & \xmark & \checkmark\\ \hline
        LR + LSTM Hybrid (+ White Noise) & LR-LSTM (-AWN) & \ref{subsec:Hybrid} & state, LR residuals & tendency & OLS, MAE & \checkmark & \checkmark\\
    \end{tabular}
    \caption{An overview of all the methods considered in this work: method, abbreviation, the section in which they are explained, input, output, cost function, presence of memory effects (in model definition/construction), and state dependency/independency of noise (only applicable for stochastic models). 
    OLS and MAE stand for Ordinary Least Square and Mean Absolute Error, respectively; NA means not applicable. }
    \label{tab:overview}
\end{table}
\end{linenomath*}

A solution of the idealized, eddy-resolving, double-gyre, quasigeostrophic ocean circulation model is used as the reference dataset.
The testbed for the reduced-order modeling consists of the leading $150$ Empirical Orthogonal Functions (EOFs) / Principal Components (PCs) of the reference solutions (out of a total $50,000$), and these capture about $90\%$ of the total variance.
The higher-ranked EOFs show multi-scale variability along the turbulent eastward jet region, while the low-ranked ones show small-scale variations in the entire domain; the corresponding PCs exhibit structured (only for the top few) and noisy patterns, referring to a mixture of low- and high-frequency variability.
We have modeled the PCs' dynamics using different methodologies, obtained forecasts, reconstructed the physical spatio-temporal fields using the EOFs, assessed the results using several assessment metrics, and inter-compared model skills.
Training of the models is done using a $\approx 1100$ years long dataset with $10$ days as the sampling period.
We have studied forecasts with both short- and long-timescales, where the lead times are on seasonal and centennial time scales, respectively.
To assess the accuracy of short-time forecasts, we have used RMSE and Anomaly Cross Correlations (ACC), and, for long-timescale forecasts, we used climatology, variance, and frequency maps.
Additionally, we have diagnosed the stability and computational costs of the methods using forecast horizon and training/prediction time complexities, respectively.
It is also possible to define a few dynamically inspired performance metrics, such as `product integral' discussed in \citeA{niraj2020eddyforcing}, for future eddy emulators, but developing and applying such metrics is beyond the scope of this paper.

\noindent We have made the following key observations during the assessment:
\begin{itemize}
    \item On short forecast lead times (e.g., several months), Multi-Level LR (ML-LR) delivers the best results, whereas the standalone deep-learning methods, both in the presence and absence of the noise, perform the worst, as evidenced by a higher RMSE and poor ACC's both in space and time.
    The hybrid methods with additive noise are the next best after ML-LR; the ANN-hybrid model performs marginally better than the LSTM-hybrid. 
    The success of ML-LR and noise-augmented hybrid models reflects the importance of memory effects and stochastic noise when accounting for dynamical interactions with the removed EOFs/PCs (i.e., beyond the rank $150$). Including these effects, accounting for the truncated dynamics, is rigorously justified in Mori--Zwanzig theory of statistical mechanics \cite{MSM2015}, which started also to attract attention in deep learning \cite{WANG2020}.
    The ML-LR conveys memory effects using an additional LR level, leading to a red-type noise, whereas LSTM achieves this by definition.
    The state dependency of noise is achieved using additional regressions in ML-LR, whereas in hybrid models this is accomplished using state variables as inputs alongside the LR residuals to forecast the residuals for the next time instant. 
    
    Additionally, both models explicitly resolve the linear dependency, which is vital because pure ANN/LSTM (with or without noise) display a performance even worse than persistence for short lead times.
    The linear dynamics is pertinent to the gyres and must explain a significant proportion of the variance in the top PCs. 
    Representing it using linear regression (i) determines the coefficients accurate up to the machine precision, (ii) leaves a lesser number of terms to be learned more, and (iii) improves the signal-to-noise ratio in the residual statistics on which ANN/LSTM is later trained to a higher effect.  
    Therefore, we argue that the use of bare ANN/LSTM is not useful in situations where linear terms dominate that can be learned via regression (similar to \citeA{pyle2021domain}).
    In such cases, it is better to use the ANN/LSTM as a correction (potentially stochastic) term in combination with LR to get an optimal closure model for the dynamics of the retained EOFs.
    
    For long-timescale forecasts (e.g., several centuries), the white-noise extension of LR, ML-LR, and stochastic hybrid models perform the best as they correctly resolve both low- and high-frequency variabilities across the domain (with the right frequency magnitude) and as they produce low climatological bias. 
    Stochastic ANN-hybrid produces a lower climatological bias than its LSTM counterpart, but the latter produces relatively better frequency map; the variance estimates are almost the same for both. 
    LR outputs decay to zero, whereas all standalone deep-learning methods (with or without noise) generate strong climatological drifts and fail to represent the flow variability both in the jet and the gyre regions.
    
    \item All models show a better forecast horizon than the LR.
    In particular, all the noise-augmented models produce stable, non-zero, and finite forecasts on climate-like lead times, e.g., centuries and millennia, meaning that adding noise to deterministic models alongside the system variables boosts its stability while keeping the forecasts bounded and physically relevant.
    
    \item Simple white noise extension of LR shows a similar training and prediction time complexity as the LR model, but they become more expensive for more complex architecture, such as the ML-LR.
    Deep-learning-based methods are the most expensive to train but relatively cheaper during forecasts -- a potentially useful property from a climate forecasting viewpoint.
    Note that LSTM models are more expensive to train and forecast than the others, so it is better to use them in the hybrid configuration, as they benefit from a simpler model configuration, and thus fewer trainable parameters and lesser training (and forecast) time.
\end{itemize}

Overall, our results prove the superiority of ML-LR and stochastically augmented hybrid models for building simple, stable, and low-cost reduced-order ocean emulators (within the EOFs/PCs framework) for producing short-/long-timescale forecasts.
The success of these methods highlights the importance of including core dynamics, memory effects, and model errors for building reliable emulators.
In this application, we have considered the core dynamics as linear and concentrated on the latter two components to prove that state-dependent additive noise is an excellent way to account for memory and unknown dynamical model errors in emulators.
ML-LR allows for only linear additive state-dependent noise (and memory), but hybrid deep-learning models can potentially learn very general forms of non-linear and multiplicative state-dependent noise. 
Similar outcomes of ML-LR and hybrid deep-learning models prove that ML-LR produces the most optimal noise configuration which deep learning learned successfully. 
Another evidence of the importance of state-dependent noise is the poor performance of a purely red noise augmentation of LR (not shown), which has memory but no state-dependency for the noise.

However, as for any data-driven method, the results presented here are valid for the current training length, and we admit that a more prolonged training may improve the models' performance.
However, using a longer training length can be computationally prohibitive, and too-long ocean observations may not be available in real life.
Therefore the emphasis here is also on identifying the models that perform relatively well even on shorter training data.
The use of orthogonal bases, i.e., EOFs/PCs, puts another constraint on the current study, as many fluid dynamical systems may not follow this assumption, but using and comparing different bases transformations, e.g., Dynamical Mode Decomposition (DMD) modes, is beyond the scope of the paper and is left as a future exercise.

The current research can be extended along the following lines. 
The first and also straightforward direction is to test the performance of the best-performing stochastic models on a more complicated testbed, e.g., on the output of a comprehensive ocean general circulation model or coupled ocean-atmosphere models for emulating, say, ENSO or Madden-Julien Oscillation, which incorporate significant delay time response . 
Such an implementation would further ascertain our findings for building reliable emulators for complicated ocean/atmospheric processes.
It is worth looking for the ways of imposing physical constraints, such as energy/mass conservation, into the emulators, e.g., using an appropriate penalizing term in the loss function.
Secondly, the results obtained here are directly relevant for emulation of various complex and multi-scale fields in the context of eddy parameterizations in low-resolution ocean models. 
Finally, a possible sequel to this work is including more stochastic and deep-learning methods, or a mixture of both, e.g., the Stochastic Neural Networks.
We started to develop a rigorous testbed for data-driven models and used this for several model setups, but we will expand this to more complex testbeds, models and datasets in the future and check if the conclusions still hold. 
\acknowledgments
The authors are thankful to the four anonymous reviewers for their feedback and suggestions, which improved the manuscript significantly. 
NA is grateful to the Research Computing Service (RCS) team of Imperial College London for the help and assistance with HPC, and to MPE CDT for providing the financial and technical support for conducting this research. 
PD gratefully acknowledges funding from the Royal Society for his University Research Fellowship as well as the ESiWACE, MAELSTROM and AI4Copernicus under Horizon 2020 and the European High-Performance Computing Joint Undertaking (JU; grant agreement No 823988, 955513 and 101016798). The JU received funding from the European High-Performance Computing Joint Undertaking (JU) under grant agreement No 955513. 
The JU receives support from the European Union's Horizon 2020 research and innovation programme and United Kingdom, Germany, Italy, Luxembourg, Switzerland, Norway.
The ESIWACE2 project has received funding from the European Union’s Horizon 2020 research and innovation program under grant agreement No. 823988.
PB and ER are supported by NERC grant NE/R011567/1  and  Royal  Society  Exchange  Grant  IEC/R2/181033,  DK  is supported by the  National Science Foundation grant OCE-1658357. Analysis in Secs.~\ref{subsec:MLLR} and \ref{subsec:Hybrid}  was supported by the Russian Science Foundation (Grant No. 18-12-00231). The part of the study conducted by ER that involved producing data with varying spatial resolutions of the baroclinic ocean flow model for further analysis was partly supported by the Russian Science Foundation (Project 19-17-00006).
PB  gratefully acknowledges the support by NERC grant NE/T002220/1 and Leverhulme grant RPG-2019-024. 
PB was also supported by the Moscow Center for Fundamental and Applied Mathematics (supported by the Agreement 075-15-2019-1624 with the Ministry of Education and Science of the Russian Federation). 

\noindent ML-LR Toolbox is available at research.atmos.ucla.edu/tcd/dkondras/Software.html. 
The source codes for the other methods and the dataset used can be found in the figshare repository: https://doi.org/10.6084/m9.figshare.14188349.v4. 
\bibliography{bibliography}

\begin{thebibliography}{}

\bibitem [\protect \citeauthoryear {%
Agarwal%
, Ryzhov%
, Kondrashov%
\BCBL {}\ \BBA {} Berloff%
}{%
Agarwal%
\ \protect \BOthers {.}}{%
{\protect \APACyear {2020}}%
}]{%
niraj2020eddyforcing}
\APACinsertmetastar {%
niraj2020eddyforcing}%
\begin{APACrefauthors}%
Agarwal, N.%
, Ryzhov, E.%
, Kondrashov, D.%
\BCBL {}\ \BBA {} Berloff, P.%
\end{APACrefauthors}%
\unskip\
\newblock
\APACrefYearMonthDay{2020}{}{}.
\newblock
{\BBOQ}\APACrefatitle {Scale-aware flow decomposition and statistical analysis
  of the eddy forcing} {Scale-aware flow decomposition and statistical analysis
  of the eddy forcing}.{\BBCQ}
\newblock
\APACjournalVolNumPages{submitted to Journal of Fluid Mechanics}{}{}{}.
\PrintBackRefs{\CurrentBib}

\bibitem [\protect \citeauthoryear {%
P.~Berloff%
}{%
P.~Berloff%
}{%
{\protect \APACyear {2015}}%
}]{%
berloff2015dynamically}
\APACinsertmetastar {%
berloff2015dynamically}%
\begin{APACrefauthors}%
Berloff, P.%
\end{APACrefauthors}%
\unskip\
\newblock
\APACrefYearMonthDay{2015}{}{}.
\newblock
{\BBOQ}\APACrefatitle {Dynamically consistent parameterization of mesoscale
  eddies. Part I: {S}imple model} {Dynamically consistent parameterization of
  mesoscale eddies. part i: {S}imple model}.{\BBCQ}
\newblock
\APACjournalVolNumPages{Ocean Modelling}{87}{}{1--19}.
\PrintBackRefs{\CurrentBib}

\bibitem [\protect \citeauthoryear {%
P\BPBI S.~Berloff%
\ \BBA {} McWilliams%
}{%
P\BPBI S.~Berloff%
\ \BBA {} McWilliams%
}{%
{\protect \APACyear {2003}}%
}]{%
berloff2003material}
\APACinsertmetastar {%
berloff2003material}%
\begin{APACrefauthors}%
Berloff, P\BPBI S.%
\BCBT {}\ \BBA {} McWilliams, J\BPBI C.%
\end{APACrefauthors}%
\unskip\
\newblock
\APACrefYearMonthDay{2003}{}{}.
\newblock
{\BBOQ}\APACrefatitle {Material transport in oceanic gyres. Part III:
  Randomized stochastic models} {Material transport in oceanic gyres. part iii:
  Randomized stochastic models}.{\BBCQ}
\newblock
\APACjournalVolNumPages{Journal of physical oceanography}{33}{7}{1416--1445}.
\PrintBackRefs{\CurrentBib}

\bibitem [\protect \citeauthoryear {%
Berner%
\ \protect \BOthers {.}}{%
Berner%
\ \protect \BOthers {.}}{%
{\protect \APACyear {2017}}%
}]{%
berner2017stochastic}
\APACinsertmetastar {%
berner2017stochastic}%
\begin{APACrefauthors}%
Berner, J.%
, Achatz, U.%
, Batte, L.%
, Bengtsson, L.%
, C{\'a}mara, A\BPBI d\BPBI l.%
, Christensen, H\BPBI M.%
\BDBL {}others%
\end{APACrefauthors}%
\unskip\
\newblock
\APACrefYearMonthDay{2017}{}{}.
\newblock
{\BBOQ}\APACrefatitle {Stochastic parameterization: Toward a new view of
  weather and climate models} {Stochastic parameterization: Toward a new view
  of weather and climate models}.{\BBCQ}
\newblock
\APACjournalVolNumPages{Bulletin of the American Meteorological
  Society}{98}{3}{565--588}.
\PrintBackRefs{\CurrentBib}

\bibitem [\protect \citeauthoryear {%
Bolton%
\ \BBA {} Zanna%
}{%
Bolton%
\ \BBA {} Zanna%
}{%
{\protect \APACyear {2019}}%
}]{%
bolton2019applications}
\APACinsertmetastar {%
bolton2019applications}%
\begin{APACrefauthors}%
Bolton, T.%
\BCBT {}\ \BBA {} Zanna, L.%
\end{APACrefauthors}%
\unskip\
\newblock
\APACrefYearMonthDay{2019}{}{}.
\newblock
{\BBOQ}\APACrefatitle {Applications of deep learning to ocean data inference
  and subgrid parameterization} {Applications of deep learning to ocean data
  inference and subgrid parameterization}.{\BBCQ}
\newblock
\APACjournalVolNumPages{Journal of Advances in Modeling Earth
  Systems}{11}{1}{376--399}.
\PrintBackRefs{\CurrentBib}

\bibitem [\protect \citeauthoryear {%
Brenner%
, Eldredge%
\BCBL {}\ \BBA {} Freund%
}{%
Brenner%
\ \protect \BOthers {.}}{%
{\protect \APACyear {2019}}%
}]{%
brenner2019perspective}
\APACinsertmetastar {%
brenner2019perspective}%
\begin{APACrefauthors}%
Brenner, M.%
, Eldredge, J.%
\BCBL {}\ \BBA {} Freund, J.%
\end{APACrefauthors}%
\unskip\
\newblock
\APACrefYearMonthDay{2019}{}{}.
\newblock
{\BBOQ}\APACrefatitle {Perspective on machine learning for advancing fluid
  mechanics} {Perspective on machine learning for advancing fluid
  mechanics}.{\BBCQ}
\newblock
\APACjournalVolNumPages{Physical Review Fluids}{4}{10}{100501}.
\PrintBackRefs{\CurrentBib}

\bibitem [\protect \citeauthoryear {%
Buizza%
, Milleer%
\BCBL {}\ \BBA {} Palmer%
}{%
Buizza%
\ \protect \BOthers {.}}{%
{\protect \APACyear {1999}}%
}]{%
buizza1999stochastic}
\APACinsertmetastar {%
buizza1999stochastic}%
\begin{APACrefauthors}%
Buizza, R.%
, Milleer, M.%
\BCBL {}\ \BBA {} Palmer, T\BPBI N.%
\end{APACrefauthors}%
\unskip\
\newblock
\APACrefYearMonthDay{1999}{}{}.
\newblock
{\BBOQ}\APACrefatitle {Stochastic representation of model uncertainties in the
  {ECMWF} ensemble prediction system} {Stochastic representation of model
  uncertainties in the {ECMWF} ensemble prediction system}.{\BBCQ}
\newblock
\APACjournalVolNumPages{Quarterly Journal of the Royal Meteorological
  Society}{125}{560}{2887--2908}.
\PrintBackRefs{\CurrentBib}

\bibitem [\protect \citeauthoryear {%
Chattopadhyay%
, Hassanzadeh%
\BCBL {}\ \BBA {} Subramanian%
}{%
Chattopadhyay%
, Hassanzadeh%
\BCBL {}\ \BBA {} Subramanian%
}{%
{\protect \APACyear {2020}}%
}]{%
chattopadhyay2020dataNP}
\APACinsertmetastar {%
chattopadhyay2020dataNP}%
\begin{APACrefauthors}%
Chattopadhyay, A.%
, Hassanzadeh, P.%
\BCBL {}\ \BBA {} Subramanian, D.%
\end{APACrefauthors}%
\unskip\
\newblock
\APACrefYearMonthDay{2020}{}{}.
\newblock
{\BBOQ}\APACrefatitle {Data-driven predictions of a multiscale Lorenz 96
  chaotic system using machine-learning methods: reservoir computing,
  artificial neural network, and long short-term memory network} {Data-driven
  predictions of a multiscale lorenz 96 chaotic system using machine-learning
  methods: reservoir computing, artificial neural network, and long short-term
  memory network}.{\BBCQ}
\newblock
\APACjournalVolNumPages{Nonlinear Processes in Geophysics}{27}{3}{373--389}.
\PrintBackRefs{\CurrentBib}

\bibitem [\protect \citeauthoryear {%
Chattopadhyay%
, Mustafa%
, Hassanzadeh%
\BCBL {}\ \BBA {} Kashinath%
}{%
Chattopadhyay%
, Mustafa%
\BCBL {}\ \protect \BOthers {.}}{%
{\protect \APACyear {2020}}%
}]{%
chattopadhyay2020deep}
\APACinsertmetastar {%
chattopadhyay2020deep}%
\begin{APACrefauthors}%
Chattopadhyay, A.%
, Mustafa, M.%
, Hassanzadeh, P.%
\BCBL {}\ \BBA {} Kashinath, K.%
\end{APACrefauthors}%
\unskip\
\newblock
\APACrefYearMonthDay{2020}{}{}.
\newblock
{\BBOQ}\APACrefatitle {Deep spatial transformers for autoregressive data-driven
  forecasting of geophysical turbulence} {Deep spatial transformers for
  autoregressive data-driven forecasting of geophysical turbulence}.{\BBCQ}
\newblock
\BIn{} \APACrefbtitle {Proceedings of the 10th International Conference on
  Climate Informatics} {Proceedings of the 10th international conference on
  climate informatics}\ (\BPGS\ 106--112).
\PrintBackRefs{\CurrentBib}

\bibitem [\protect \citeauthoryear {%
Chattopadhyay%
, Subel%
\BCBL {}\ \BBA {} Hassanzadeh%
}{%
Chattopadhyay%
, Subel%
\BCBL {}\ \BBA {} Hassanzadeh%
}{%
{\protect \APACyear {2020}}%
}]{%
chattopadhyay2020data}
\APACinsertmetastar {%
chattopadhyay2020data}%
\begin{APACrefauthors}%
Chattopadhyay, A.%
, Subel, A.%
\BCBL {}\ \BBA {} Hassanzadeh, P.%
\end{APACrefauthors}%
\unskip\
\newblock
\APACrefYearMonthDay{2020}{}{}.
\newblock
{\BBOQ}\APACrefatitle {Data-Driven Super-Parameterization Using Deep Learning:
  Experimentation With Multiscale Lorenz 96 Systems and Transfer Learning}
  {Data-driven super-parameterization using deep learning: Experimentation with
  multiscale lorenz 96 systems and transfer learning}.{\BBCQ}
\newblock
\APACjournalVolNumPages{Journal of Advances in Modeling Earth
  Systems}{12}{11}{e2020MS002084}.
\PrintBackRefs{\CurrentBib}

\bibitem [\protect \citeauthoryear {%
Chen%
\ \protect \BOthers {.}}{%
Chen%
\ \protect \BOthers {.}}{%
{\protect \APACyear {2016}}%
}]{%
Chen_etal2016}
\APACinsertmetastar {%
Chen_etal2016}%
\begin{APACrefauthors}%
Chen, C.%
, Cane, M\BPBI A.%
, Henderson, N.%
, Lee, D\BPBI E.%
, Chapman, D.%
, Kondrashov, D.%
\BCBL {}\ \BBA {} Chekroun, M\BPBI D.%
\end{APACrefauthors}%
\unskip\
\newblock
\APACrefYearMonthDay{2016}{}{}.
\newblock
{\BBOQ}\APACrefatitle {{Diversity, Nonlinearity, Seasonality, and Memory Effect
  in {ENSO} Simulation and Prediction Using Empirical Model Reduction}}
  {{Diversity, Nonlinearity, Seasonality, and Memory Effect in {ENSO}
  Simulation and Prediction Using Empirical Model Reduction}}.{\BBCQ}
\newblock
\APACjournalVolNumPages{Journal of Climate}{29}{5}{1809-1830}.
\newblock
\begin{APACrefDOI} \doi{10.1175/JCLI-D-15-0372.1} \end{APACrefDOI}
\PrintBackRefs{\CurrentBib}

\bibitem [\protect \citeauthoryear {%
D'Andrea%
\ \BBA {} Vautard%
}{%
D'Andrea%
\ \BBA {} Vautard%
}{%
{\protect \APACyear {2001}}%
}]{%
d2001extratropical}
\APACinsertmetastar {%
d2001extratropical}%
\begin{APACrefauthors}%
D'Andrea, F.%
\BCBT {}\ \BBA {} Vautard, R.%
\end{APACrefauthors}%
\unskip\
\newblock
\APACrefYearMonthDay{2001}{}{}.
\newblock
{\BBOQ}\APACrefatitle {Extratropical low-frequency variability as a
  low-dimensional problem I: A simplified model} {Extratropical low-frequency
  variability as a low-dimensional problem i: A simplified model}.{\BBCQ}
\newblock
\APACjournalVolNumPages{Quarterly Journal of the Royal Meteorological
  Society}{127}{574}{1357--1374}.
\PrintBackRefs{\CurrentBib}

\bibitem [\protect \citeauthoryear {%
Debussche%
, Glatt-Holtz%
, Temam%
\BCBL {}\ \BBA {} Ziane%
}{%
Debussche%
\ \protect \BOthers {.}}{%
{\protect \APACyear {2012}}%
}]{%
debussche2012global}
\APACinsertmetastar {%
debussche2012global}%
\begin{APACrefauthors}%
Debussche, A.%
, Glatt-Holtz, N.%
, Temam, R.%
\BCBL {}\ \BBA {} Ziane, M.%
\end{APACrefauthors}%
\unskip\
\newblock
\APACrefYearMonthDay{2012}{}{}.
\newblock
{\BBOQ}\APACrefatitle {Global existence and regularity for the 3{D} stochastic
  primitive equations of the ocean and atmosphere with multiplicative white
  noise} {Global existence and regularity for the 3{D} stochastic primitive
  equations of the ocean and atmosphere with multiplicative white
  noise}.{\BBCQ}
\newblock
\APACjournalVolNumPages{Nonlinearity}{25}{7}{2093}.
\PrintBackRefs{\CurrentBib}

\bibitem [\protect \citeauthoryear {%
DelSole%
}{%
DelSole%
}{%
{\protect \APACyear {2004}}%
}]{%
delsole2004stochastic}
\APACinsertmetastar {%
delsole2004stochastic}%
\begin{APACrefauthors}%
DelSole, T.%
\end{APACrefauthors}%
\unskip\
\newblock
\APACrefYearMonthDay{2004}{}{}.
\newblock
{\BBOQ}\APACrefatitle {Stochastic models of quasigeostrophic turbulence}
  {Stochastic models of quasigeostrophic turbulence}.{\BBCQ}
\newblock
\APACjournalVolNumPages{Surveys in Geophysics}{25}{2}{107--149}.
\PrintBackRefs{\CurrentBib}

\bibitem [\protect \citeauthoryear {%
DelSole%
\ \BBA {} Hou%
}{%
DelSole%
\ \BBA {} Hou%
}{%
{\protect \APACyear {1999}}%
}]{%
delsole1999empirical}
\APACinsertmetastar {%
delsole1999empirical}%
\begin{APACrefauthors}%
DelSole, T.%
\BCBT {}\ \BBA {} Hou, A\BPBI Y.%
\end{APACrefauthors}%
\unskip\
\newblock
\APACrefYearMonthDay{1999}{}{}.
\newblock
{\BBOQ}\APACrefatitle {Empirical stochastic models for the dominant climate
  statistics of a general circulation model} {Empirical stochastic models for
  the dominant climate statistics of a general circulation model}.{\BBCQ}
\newblock
\APACjournalVolNumPages{Journal of the atmospheric
  sciences}{56}{19}{3436--3456}.
\PrintBackRefs{\CurrentBib}

\bibitem [\protect \citeauthoryear {%
Dueben%
\ \BBA {} Bauer%
}{%
Dueben%
\ \BBA {} Bauer%
}{%
{\protect \APACyear {2018}}%
}]{%
gmd-11-3999-2018}
\APACinsertmetastar {%
gmd-11-3999-2018}%
\begin{APACrefauthors}%
Dueben, P\BPBI D.%
\BCBT {}\ \BBA {} Bauer, P.%
\end{APACrefauthors}%
\unskip\
\newblock
\APACrefYearMonthDay{2018}{}{}.
\newblock
{\BBOQ}\APACrefatitle {Challenges and design choices for global weather and
  climate models based on machine learning} {Challenges and design choices for
  global weather and climate models based on machine learning}.{\BBCQ}
\newblock
\APACjournalVolNumPages{Geoscientific Model Development}{11}{10}{3999--4009}.
\newblock
\begin{APACrefURL} \url{https://www.geosci-model-dev.net/11/3999/2018/}
  \end{APACrefURL}
\newblock
\begin{APACrefDOI} \doi{10.5194/gmd-11-3999-2018} \end{APACrefDOI}
\PrintBackRefs{\CurrentBib}

\bibitem [\protect \citeauthoryear {%
Ewald%
, Petcu%
\BCBL {}\ \BBA {} Temam%
}{%
Ewald%
\ \protect \BOthers {.}}{%
{\protect \APACyear {2007}}%
}]{%
ewald2007stochastic}
\APACinsertmetastar {%
ewald2007stochastic}%
\begin{APACrefauthors}%
Ewald, B.%
, Petcu, M.%
\BCBL {}\ \BBA {} Temam, R.%
\end{APACrefauthors}%
\unskip\
\newblock
\APACrefYearMonthDay{2007}{}{}.
\newblock
{\BBOQ}\APACrefatitle {Stochastic solutions of the two-dimensional primitive
  equations of the ocean and atmosphere with an additive noise} {Stochastic
  solutions of the two-dimensional primitive equations of the ocean and
  atmosphere with an additive noise}.{\BBCQ}
\newblock
\APACjournalVolNumPages{Analysis and Applications}{5}{02}{183--198}.
\PrintBackRefs{\CurrentBib}

\bibitem [\protect \citeauthoryear {%
Farrell%
\ \BBA {} Ioannou%
}{%
Farrell%
\ \BBA {} Ioannou%
}{%
{\protect \APACyear {1993}}%
}]{%
farrell1993stochastic}
\APACinsertmetastar {%
farrell1993stochastic}%
\begin{APACrefauthors}%
Farrell, B\BPBI F.%
\BCBT {}\ \BBA {} Ioannou, P\BPBI J.%
\end{APACrefauthors}%
\unskip\
\newblock
\APACrefYearMonthDay{1993}{}{}.
\newblock
{\BBOQ}\APACrefatitle {Stochastic forcing of the linearized Navier--Stokes
  equations} {Stochastic forcing of the linearized navier--stokes
  equations}.{\BBCQ}
\newblock
\APACjournalVolNumPages{Physics of Fluids A: Fluid
  Dynamics}{5}{11}{2600--2609}.
\PrintBackRefs{\CurrentBib}

\bibitem [\protect \citeauthoryear {%
Farrell%
\ \BBA {} Ioannou%
}{%
Farrell%
\ \BBA {} Ioannou%
}{%
{\protect \APACyear {1995}}%
}]{%
farrell1995stochastic}
\APACinsertmetastar {%
farrell1995stochastic}%
\begin{APACrefauthors}%
Farrell, B\BPBI F.%
\BCBT {}\ \BBA {} Ioannou, P\BPBI J.%
\end{APACrefauthors}%
\unskip\
\newblock
\APACrefYearMonthDay{1995}{}{}.
\newblock
{\BBOQ}\APACrefatitle {Stochastic dynamics of the midlatitude atmospheric jet}
  {Stochastic dynamics of the midlatitude atmospheric jet}.{\BBCQ}
\newblock
\APACjournalVolNumPages{Journal of the atmospheric
  sciences}{52}{10}{1642--1656}.
\PrintBackRefs{\CurrentBib}

\bibitem [\protect \citeauthoryear {%
C.~Franzke%
}{%
C.~Franzke%
}{%
{\protect \APACyear {2012}}%
}]{%
franzke2012predictability}
\APACinsertmetastar {%
franzke2012predictability}%
\begin{APACrefauthors}%
Franzke, C.%
\end{APACrefauthors}%
\unskip\
\newblock
\APACrefYearMonthDay{2012}{}{}.
\newblock
{\BBOQ}\APACrefatitle {Predictability of extreme events in a nonlinear
  stochastic-dynamical model} {Predictability of extreme events in a nonlinear
  stochastic-dynamical model}.{\BBCQ}
\newblock
\APACjournalVolNumPages{Physical Review E}{85}{3}{031134}.
\PrintBackRefs{\CurrentBib}

\bibitem [\protect \citeauthoryear {%
C.~Franzke%
, Majda%
\BCBL {}\ \BBA {} Vanden-Eijnden%
}{%
C.~Franzke%
\ \protect \BOthers {.}}{%
{\protect \APACyear {2005}}%
}]{%
franzke2005low}
\APACinsertmetastar {%
franzke2005low}%
\begin{APACrefauthors}%
Franzke, C.%
, Majda, A\BPBI J.%
\BCBL {}\ \BBA {} Vanden-Eijnden, E.%
\end{APACrefauthors}%
\unskip\
\newblock
\APACrefYearMonthDay{2005}{}{}.
\newblock
{\BBOQ}\APACrefatitle {Low-order stochastic mode reduction for a realistic
  barotropic model climate} {Low-order stochastic mode reduction for a
  realistic barotropic model climate}.{\BBCQ}
\newblock
\APACjournalVolNumPages{Journal of the atmospheric
  sciences}{62}{6}{1722--1745}.
\PrintBackRefs{\CurrentBib}

\bibitem [\protect \citeauthoryear {%
C\BPBI L.~Franzke%
}{%
C\BPBI L.~Franzke%
}{%
{\protect \APACyear {2013}}%
}]{%
franzke2013predictions}
\APACinsertmetastar {%
franzke2013predictions}%
\begin{APACrefauthors}%
Franzke, C\BPBI L.%
\end{APACrefauthors}%
\unskip\
\newblock
\APACrefYearMonthDay{2013}{}{}.
\newblock
{\BBOQ}\APACrefatitle {Predictions of critical transitions with non-stationary
  reduced order models} {Predictions of critical transitions with
  non-stationary reduced order models}.{\BBCQ}
\newblock
\APACjournalVolNumPages{Physica D: Nonlinear Phenomena}{262}{}{35--47}.
\PrintBackRefs{\CurrentBib}

\bibitem [\protect \citeauthoryear {%
Ghil%
\ \protect \BOthers {.}}{%
Ghil%
\ \protect \BOthers {.}}{%
{\protect \APACyear {2002}}%
}]{%
Ghil.review.ea.2002}
\APACinsertmetastar {%
Ghil.review.ea.2002}%
\begin{APACrefauthors}%
Ghil, M.%
, Allen, M\BPBI R.%
, Dettinger, M\BPBI D.%
, Ide, K.%
, Kondrashov, D.%
\BCBL {}\ \BOthersPeriod {.}\end{APACrefauthors}%
\unskip\
\newblock
\APACrefYearMonthDay{2002}{}{}.
\newblock
{\BBOQ}\APACrefatitle {Advanced spectral methods for climatic time series}
  {Advanced spectral methods for climatic time series}.{\BBCQ}
\newblock
\APACjournalVolNumPages{Review of Geophysics}{40}{1}{1--41}.
\PrintBackRefs{\CurrentBib}

\bibitem [\protect \citeauthoryear {%
Ghil%
, Groth%
, Kondrashov%
\BCBL {}\ \BBA {} Robertson%
}{%
Ghil%
\ \protect \BOthers {.}}{%
{\protect \APACyear {2018}}%
}]{%
GGKR18}
\APACinsertmetastar {%
GGKR18}%
\begin{APACrefauthors}%
Ghil, M.%
, Groth, A.%
, Kondrashov, D.%
\BCBL {}\ \BBA {} Robertson, A.%
\end{APACrefauthors}%
\unskip\
\newblock
\APACrefYearMonthDay{2018}{}{}.
\newblock
{\BBOQ}\APACrefatitle {{Extratropical sub-seasonal--to--seasonal oscillations
  and multiple regimes: The dynamical systems view}} {{Extratropical
  sub-seasonal--to--seasonal oscillations and multiple regimes: The dynamical
  systems view}}.{\BBCQ}
\newblock
\BIn{} { A. W. Robertson and F. Vitart}\ (\BED), \APACrefbtitle {The Gap
  between Weather and Climate Forecasting: Sub-Seasonal to Seasonal Prediction}
  {The gap between weather and climate forecasting: Sub-seasonal to seasonal
  prediction}\ (\BPGS\ 119--142).
\newblock
\APACaddressPublisher{}{Elsevier}.
\PrintBackRefs{\CurrentBib}

\bibitem [\protect \citeauthoryear {%
Glatt-Holtz%
\ \BBA {} Ziane%
}{%
Glatt-Holtz%
\ \BBA {} Ziane%
}{%
{\protect \APACyear {2008}}%
}]{%
glatt2008stochastic}
\APACinsertmetastar {%
glatt2008stochastic}%
\begin{APACrefauthors}%
Glatt-Holtz, N.%
\BCBT {}\ \BBA {} Ziane, M.%
\end{APACrefauthors}%
\unskip\
\newblock
\APACrefYearMonthDay{2008}{}{}.
\newblock
{\BBOQ}\APACrefatitle {The stochastic primitive equations in two space
  dimensions with multiplicative noise} {The stochastic primitive equations in
  two space dimensions with multiplicative noise}.{\BBCQ}
\newblock
\APACjournalVolNumPages{Discrete Contin. Dyn. Syst. Ser. B}{10}{4}{801--822}.
\PrintBackRefs{\CurrentBib}

\bibitem [\protect \citeauthoryear {%
Ham%
, Kim%
\BCBL {}\ \BBA {} Luo%
}{%
Ham%
\ \protect \BOthers {.}}{%
{\protect \APACyear {2019}}%
}]{%
ham2019deep}
\APACinsertmetastar {%
ham2019deep}%
\begin{APACrefauthors}%
Ham, Y\BHBI G.%
, Kim, J\BHBI H.%
\BCBL {}\ \BBA {} Luo, J\BHBI J.%
\end{APACrefauthors}%
\unskip\
\newblock
\APACrefYearMonthDay{2019}{}{}.
\newblock
{\BBOQ}\APACrefatitle {Deep learning for multi-year ENSO forecasts} {Deep
  learning for multi-year enso forecasts}.{\BBCQ}
\newblock
\APACjournalVolNumPages{Nature}{573}{7775}{568--572}.
\PrintBackRefs{\CurrentBib}

\bibitem [\protect \citeauthoryear {%
Hochreiter%
\ \BBA {} Schmidhuber%
}{%
Hochreiter%
\ \BBA {} Schmidhuber%
}{%
{\protect \APACyear {1997}}%
}]{%
hochreiter1997long}
\APACinsertmetastar {%
hochreiter1997long}%
\begin{APACrefauthors}%
Hochreiter, S.%
\BCBT {}\ \BBA {} Schmidhuber, J.%
\end{APACrefauthors}%
\unskip\
\newblock
\APACrefYearMonthDay{1997}{}{}.
\newblock
{\BBOQ}\APACrefatitle {Long short-term memory} {Long short-term memory}.{\BBCQ}
\newblock
\APACjournalVolNumPages{Neural computation}{9}{8}{1735--1780}.
\PrintBackRefs{\CurrentBib}

\bibitem [\protect \citeauthoryear {%
P.~Holden%
\ \protect \BOthers {.}}{%
P.~Holden%
\ \protect \BOthers {.}}{%
{\protect \APACyear {2013}}%
}]{%
holden2013plasim}
\APACinsertmetastar {%
holden2013plasim}%
\begin{APACrefauthors}%
Holden, P.%
, Edwards, N.%
, Garthwaite, P.%
, Fraedrich, K.%
, Lunkeit, F.%
, Kirk, E.%
\BDBL {}Babonneau, F.%
\end{APACrefauthors}%
\unskip\
\newblock
\APACrefYearMonthDay{2013}{}{}.
\newblock
{\BBOQ}\APACrefatitle {{PLASIM-ENTS}em: a spatio-temporal emulator of future
  climate change for impacts assessment} {{PLASIM-ENTS}em: a spatio-temporal
  emulator of future climate change for impacts assessment}.{\BBCQ}
\newblock
\APACjournalVolNumPages{Geoscientific model development
  discussions}{6}{2}{3349--3380}.
\PrintBackRefs{\CurrentBib}

\bibitem [\protect \citeauthoryear {%
P\BPBI B.~Holden%
, Edwards%
, Garthwaite%
\BCBL {}\ \BBA {} Wilkinson%
}{%
P\BPBI B.~Holden%
\ \protect \BOthers {.}}{%
{\protect \APACyear {2015}}%
}]{%
holden2015emulation}
\APACinsertmetastar {%
holden2015emulation}%
\begin{APACrefauthors}%
Holden, P\BPBI B.%
, Edwards, N\BPBI R.%
, Garthwaite, P\BPBI H.%
\BCBL {}\ \BBA {} Wilkinson, R\BPBI D.%
\end{APACrefauthors}%
\unskip\
\newblock
\APACrefYearMonthDay{2015}{}{}.
\newblock
{\BBOQ}\APACrefatitle {Emulation and interpretation of high-dimensional climate
  model outputs} {Emulation and interpretation of high-dimensional climate
  model outputs}.{\BBCQ}
\newblock
\APACjournalVolNumPages{Journal of Applied Statistics}{42}{9}{2038--2055}.
\PrintBackRefs{\CurrentBib}

\bibitem [\protect \citeauthoryear {%
Jaeger%
\ \BBA {} Haas%
}{%
Jaeger%
\ \BBA {} Haas%
}{%
{\protect \APACyear {2004}}%
}]{%
jaeger2004harnessing}
\APACinsertmetastar {%
jaeger2004harnessing}%
\begin{APACrefauthors}%
Jaeger, H.%
\BCBT {}\ \BBA {} Haas, H.%
\end{APACrefauthors}%
\unskip\
\newblock
\APACrefYearMonthDay{2004}{}{}.
\newblock
{\BBOQ}\APACrefatitle {Harnessing nonlinearity: Predicting chaotic systems and
  saving energy in wireless communication} {Harnessing nonlinearity: Predicting
  chaotic systems and saving energy in wireless communication}.{\BBCQ}
\newblock
\APACjournalVolNumPages{science}{304}{5667}{78--80}.
\PrintBackRefs{\CurrentBib}

\bibitem [\protect \citeauthoryear {%
Jia%
\ \protect \BOthers {.}}{%
Jia%
\ \protect \BOthers {.}}{%
{\protect \APACyear {2019}}%
}]{%
jia2019physics}
\APACinsertmetastar {%
jia2019physics}%
\begin{APACrefauthors}%
Jia, X.%
, Willard, J.%
, Karpatne, A.%
, Read, J.%
, Zwart, J.%
, Steinbach, M.%
\BCBL {}\ \BBA {} Kumar, V.%
\end{APACrefauthors}%
\unskip\
\newblock
\APACrefYearMonthDay{2019}{}{}.
\newblock
{\BBOQ}\APACrefatitle {Physics guided RNNs for modeling dynamical systems: A
  case study in simulating lake temperature profiles} {Physics guided rnns for
  modeling dynamical systems: A case study in simulating lake temperature
  profiles}.{\BBCQ}
\newblock
\BIn{} \APACrefbtitle {Proceedings of the 2019 SIAM International Conference on
  Data Mining} {Proceedings of the 2019 siam international conference on data
  mining}\ (\BPGS\ 558--566).
\PrintBackRefs{\CurrentBib}

\bibitem [\protect \citeauthoryear {%
Juricke%
, Lemke%
, Timmermann%
\BCBL {}\ \BBA {} Rackow%
}{%
Juricke%
\ \protect \BOthers {.}}{%
{\protect \APACyear {2013}}%
}]{%
juricke2013effects}
\APACinsertmetastar {%
juricke2013effects}%
\begin{APACrefauthors}%
Juricke, S.%
, Lemke, P.%
, Timmermann, R.%
\BCBL {}\ \BBA {} Rackow, T.%
\end{APACrefauthors}%
\unskip\
\newblock
\APACrefYearMonthDay{2013}{}{}.
\newblock
{\BBOQ}\APACrefatitle {Effects of stochastic ice strength perturbation on
  Arctic finite element sea ice modeling} {Effects of stochastic ice strength
  perturbation on arctic finite element sea ice modeling}.{\BBCQ}
\newblock
\APACjournalVolNumPages{Journal of climate}{26}{11}{3785--3802}.
\PrintBackRefs{\CurrentBib}

\bibitem [\protect \citeauthoryear {%
Juricke%
, Palmer%
\BCBL {}\ \BBA {} Zanna%
}{%
Juricke%
\ \protect \BOthers {.}}{%
{\protect \APACyear {2017}}%
}]{%
juricke2017stochastic}
\APACinsertmetastar {%
juricke2017stochastic}%
\begin{APACrefauthors}%
Juricke, S.%
, Palmer, T\BPBI N.%
\BCBL {}\ \BBA {} Zanna, L.%
\end{APACrefauthors}%
\unskip\
\newblock
\APACrefYearMonthDay{2017}{}{}.
\newblock
{\BBOQ}\APACrefatitle {Stochastic subgrid-scale ocean mixing: impacts on
  low-frequency variability} {Stochastic subgrid-scale ocean mixing: impacts on
  low-frequency variability}.{\BBCQ}
\newblock
\APACjournalVolNumPages{Journal of Climate}{30}{13}{4997--5019}.
\PrintBackRefs{\CurrentBib}

\bibitem [\protect \citeauthoryear {%
Karpatne%
, Atluri%
\BCBL {}\ \protect \BOthers {.}}{%
Karpatne%
, Atluri%
\BCBL {}\ \protect \BOthers {.}}{%
{\protect \APACyear {2017}}%
}]{%
karpatne2017theory}
\APACinsertmetastar {%
karpatne2017theory}%
\begin{APACrefauthors}%
Karpatne, A.%
, Atluri, G.%
, Faghmous, J\BPBI H.%
, Steinbach, M.%
, Banerjee, A.%
, Ganguly, A.%
\BDBL {}Kumar, V.%
\end{APACrefauthors}%
\unskip\
\newblock
\APACrefYearMonthDay{2017}{}{}.
\newblock
{\BBOQ}\APACrefatitle {Theory-guided data science: A new paradigm for
  scientific discovery from data} {Theory-guided data science: A new paradigm
  for scientific discovery from data}.{\BBCQ}
\newblock
\APACjournalVolNumPages{IEEE Transactions on Knowledge and Data
  Engineering}{29}{10}{2318--2331}.
\PrintBackRefs{\CurrentBib}

\bibitem [\protect \citeauthoryear {%
Karpatne%
, Watkins%
, Read%
\BCBL {}\ \BBA {} Kumar%
}{%
Karpatne%
, Watkins%
\BCBL {}\ \protect \BOthers {.}}{%
{\protect \APACyear {2017}}%
}]{%
karpatne2017physics}
\APACinsertmetastar {%
karpatne2017physics}%
\begin{APACrefauthors}%
Karpatne, A.%
, Watkins, W.%
, Read, J.%
\BCBL {}\ \BBA {} Kumar, V.%
\end{APACrefauthors}%
\unskip\
\newblock
\APACrefYearMonthDay{2017}{}{}.
\newblock
{\BBOQ}\APACrefatitle {Physics-guided neural networks (pgnn): An application in
  lake temperature modeling} {Physics-guided neural networks (pgnn): An
  application in lake temperature modeling}.{\BBCQ}
\newblock
\APACjournalVolNumPages{arXiv preprint arXiv:1710.11431}{}{}{}.
\PrintBackRefs{\CurrentBib}

\bibitem [\protect \citeauthoryear {%
Karunasinghe%
\ \BBA {} Liong%
}{%
Karunasinghe%
\ \BBA {} Liong%
}{%
{\protect \APACyear {2006}}%
}]{%
karunasinghe2006chaotic}
\APACinsertmetastar {%
karunasinghe2006chaotic}%
\begin{APACrefauthors}%
Karunasinghe, D\BPBI S.%
\BCBT {}\ \BBA {} Liong, S\BHBI Y.%
\end{APACrefauthors}%
\unskip\
\newblock
\APACrefYearMonthDay{2006}{}{}.
\newblock
{\BBOQ}\APACrefatitle {Chaotic time series prediction with a global model:
  {A}rtificial neural network} {Chaotic time series prediction with a global
  model: {A}rtificial neural network}.{\BBCQ}
\newblock
\APACjournalVolNumPages{Journal of Hydrology}{323}{1-4}{92--105}.
\PrintBackRefs{\CurrentBib}

\bibitem [\protect \citeauthoryear {%
Kondrashov%
\ \BBA {} Berloff%
}{%
Kondrashov%
\ \BBA {} Berloff%
}{%
{\protect \APACyear {2015}}%
}]{%
kondrashov2015gyres}
\APACinsertmetastar {%
kondrashov2015gyres}%
\begin{APACrefauthors}%
Kondrashov, D.%
\BCBT {}\ \BBA {} Berloff, P.%
\end{APACrefauthors}%
\unskip\
\newblock
\APACrefYearMonthDay{2015}{03}{}.
\newblock
{\BBOQ}\APACrefatitle {Stochastic modeling of decadal variability in ocean
  gyres} {Stochastic modeling of decadal variability in ocean gyres}.{\BBCQ}
\newblock
\APACjournalVolNumPages{Geophysical Research Letters}{42}{}{}.
\newblock
\begin{APACrefDOI} \doi{10.1002/2014GL062871} \end{APACrefDOI}
\PrintBackRefs{\CurrentBib}

\bibitem [\protect \citeauthoryear {%
Kondrashov%
, Chekroun%
\BCBL {}\ \BBA {} Berloff%
}{%
Kondrashov%
\ \protect \BOthers {.}}{%
{\protect \APACyear {2018}}%
}]{%
kondrashov2018multiscale}
\APACinsertmetastar {%
kondrashov2018multiscale}%
\begin{APACrefauthors}%
Kondrashov, D.%
, Chekroun, M.%
\BCBL {}\ \BBA {} Berloff, P.%
\end{APACrefauthors}%
\unskip\
\newblock
\APACrefYearMonthDay{2018}{}{}.
\newblock
{\BBOQ}\APACrefatitle {Multiscale Stuart-Landau emulators: Application to
  wind-driven ocean gyres} {Multiscale stuart-landau emulators: Application to
  wind-driven ocean gyres}.{\BBCQ}
\newblock
\APACjournalVolNumPages{Fluids}{3}{1}{21}.
\PrintBackRefs{\CurrentBib}

\bibitem [\protect \citeauthoryear {%
Kondrashov%
, Chekroun%
\BCBL {}\ \BBA {} Ghil%
}{%
Kondrashov%
\ \protect \BOthers {.}}{%
{\protect \APACyear {2015}}%
}]{%
MSM2015}
\APACinsertmetastar {%
MSM2015}%
\begin{APACrefauthors}%
Kondrashov, D.%
, Chekroun, M\BPBI D.%
\BCBL {}\ \BBA {} Ghil, M.%
\end{APACrefauthors}%
\unskip\
\newblock
\APACrefYearMonthDay{2015}{}{}.
\newblock
{\BBOQ}\APACrefatitle {Data-driven non-{Markov}ian closure models} {Data-driven
  non-{Markov}ian closure models}.{\BBCQ}
\newblock
\APACjournalVolNumPages{Physica D}{297}{}{33--55}.
\newblock
\begin{APACrefDOI} \doi{10.1016/j.physd.2014.12.005} \end{APACrefDOI}
\PrintBackRefs{\CurrentBib}

\bibitem [\protect \citeauthoryear {%
Kondrashov%
, Chekroun%
, Robertson%
\BCBL {}\ \BBA {} Ghil%
}{%
Kondrashov%
\ \protect \BOthers {.}}{%
{\protect \APACyear {2013}}%
}]{%
KCRG13}
\APACinsertmetastar {%
KCRG13}%
\begin{APACrefauthors}%
Kondrashov, D.%
, Chekroun, M\BPBI D.%
, Robertson, A\BPBI W.%
\BCBL {}\ \BBA {} Ghil, M.%
\end{APACrefauthors}%
\unskip\
\newblock
\APACrefYearMonthDay{2013}{}{}.
\newblock
{\BBOQ}\APACrefatitle {Low-order stochastic model and ``past-noise forecasting"
  of the {M}adden-{J}ulian oscillation} {Low-order stochastic model and
  ``past-noise forecasting" of the {M}adden-{J}ulian oscillation}.{\BBCQ}
\newblock
\APACjournalVolNumPages{Geophysical Research Letters}{40}{}{5305--5310}.
\PrintBackRefs{\CurrentBib}

\bibitem [\protect \citeauthoryear {%
Kondrashov%
, Kravtsov%
, Robertson%
\BCBL {}\ \BBA {} Ghil%
}{%
Kondrashov%
\ \protect \BOthers {.}}{%
{\protect \APACyear {2005}}%
}]{%
kondrashov2005a}
\APACinsertmetastar {%
kondrashov2005a}%
\begin{APACrefauthors}%
Kondrashov, D.%
, Kravtsov, S.%
, Robertson, A\BPBI W.%
\BCBL {}\ \BBA {} Ghil, M.%
\end{APACrefauthors}%
\unskip\
\newblock
\APACrefYearMonthDay{2005}{}{}.
\newblock
{\BBOQ}\APACrefatitle {A hierarchy of data-based {ENSO} models} {A hierarchy of
  data-based {ENSO} models}.{\BBCQ}
\newblock
\APACjournalVolNumPages{Journal of Climate}{18}{21}{4425--4444}.
\PrintBackRefs{\CurrentBib}

\bibitem [\protect \citeauthoryear {%
Kondrashov%
, Ryzhov%
\BCBL {}\ \BBA {} Berloff%
}{%
Kondrashov%
\ \protect \BOthers {.}}{%
{\protect \APACyear {2020}}%
}]{%
KRB2020}
\APACinsertmetastar {%
KRB2020}%
\begin{APACrefauthors}%
Kondrashov, D.%
, Ryzhov, E.%
\BCBL {}\ \BBA {} Berloff, P.%
\end{APACrefauthors}%
\unskip\
\newblock
\APACrefYearMonthDay{2020}{}{}.
\newblock
{\BBOQ}\APACrefatitle {Data-adaptive harmonic analysis of oceanic waves and
  turbulent flows} {Data-adaptive harmonic analysis of oceanic waves and
  turbulent flows}.{\BBCQ}
\newblock
\APACjournalVolNumPages{Chaos}{30}{}{061105}.
\newblock
\begin{APACrefDOI} \doi{10.1063/5.0012077} \end{APACrefDOI}
\PrintBackRefs{\CurrentBib}

\bibitem [\protect \citeauthoryear {%
Krasnopolsky%
\ \BBA {} Fox-Rabinovitz%
}{%
Krasnopolsky%
\ \BBA {} Fox-Rabinovitz%
}{%
{\protect \APACyear {2006}}%
}]{%
krasnopolsky2006complex}
\APACinsertmetastar {%
krasnopolsky2006complex}%
\begin{APACrefauthors}%
Krasnopolsky, V\BPBI M.%
\BCBT {}\ \BBA {} Fox-Rabinovitz, M\BPBI S.%
\end{APACrefauthors}%
\unskip\
\newblock
\APACrefYearMonthDay{2006}{}{}.
\newblock
{\BBOQ}\APACrefatitle {Complex hybrid models combining deterministic and
  machine learning components for numerical climate modeling and weather
  prediction} {Complex hybrid models combining deterministic and machine
  learning components for numerical climate modeling and weather
  prediction}.{\BBCQ}
\newblock
\APACjournalVolNumPages{Neural Networks}{19}{2}{122--134}.
\PrintBackRefs{\CurrentBib}

\bibitem [\protect \citeauthoryear {%
Kravtsov%
, Kondrashov%
\BCBL {}\ \BBA {} Ghil%
}{%
Kravtsov%
\ \protect \BOthers {.}}{%
{\protect \APACyear {2005}}%
}]{%
KravtsovKondrashovGhil_JCL05}
\APACinsertmetastar {%
KravtsovKondrashovGhil_JCL05}%
\begin{APACrefauthors}%
Kravtsov, S.%
, Kondrashov, D.%
\BCBL {}\ \BBA {} Ghil, M.%
\end{APACrefauthors}%
\unskip\
\newblock
\APACrefYearMonthDay{2005}{}{}.
\newblock
{\BBOQ}\APACrefatitle {Multi-level regression modeling of nonlinear processes:
  {D}erivation and applications to climatic variability} {Multi-level
  regression modeling of nonlinear processes: {D}erivation and applications to
  climatic variability}.{\BBCQ}
\newblock
\APACjournalVolNumPages{Journal of Climate}{18}{21}{4404--4424}.
\PrintBackRefs{\CurrentBib}

\bibitem [\protect \citeauthoryear {%
Kravtsov%
, Kondrashov%
\BCBL {}\ \BBA {} Ghil%
}{%
Kravtsov%
\ \protect \BOthers {.}}{%
{\protect \APACyear {2009}}%
}]{%
KravtsovGhilKondrashov_09}
\APACinsertmetastar {%
KravtsovGhilKondrashov_09}%
\begin{APACrefauthors}%
Kravtsov, S.%
, Kondrashov, D.%
\BCBL {}\ \BBA {} Ghil, M.%
\end{APACrefauthors}%
\unskip\
\newblock
\APACrefYearMonthDay{2009}{}{}.
\newblock
{\BBOQ}\APACrefatitle {Empirical Model Reduction and the Modeling Hierarchy in
  Climate Dynamics and the Geosciences} {Empirical model reduction and the
  modeling hierarchy in climate dynamics and the geosciences}.{\BBCQ}
\newblock
\BIn{} T\BPBI N.~Palmer\ \BBA {} P.~Williams\ (\BEDS), \APACrefbtitle
  {Stochastic Physics and Climate Modeling} {Stochastic physics and climate
  modeling}\ (\BPGS\ 35--72).
\newblock
\APACaddressPublisher{}{Cambridge University Press}.
\PrintBackRefs{\CurrentBib}

\bibitem [\protect \citeauthoryear {%
Krizhevsky%
, Sutskever%
\BCBL {}\ \BBA {} Hinton%
}{%
Krizhevsky%
\ \protect \BOthers {.}}{%
{\protect \APACyear {2012}}%
}]{%
krizhevsky2012imagenet}
\APACinsertmetastar {%
krizhevsky2012imagenet}%
\begin{APACrefauthors}%
Krizhevsky, A.%
, Sutskever, I.%
\BCBL {}\ \BBA {} Hinton, G\BPBI E.%
\end{APACrefauthors}%
\unskip\
\newblock
\APACrefYearMonthDay{2012}{}{}.
\newblock
{\BBOQ}\APACrefatitle {Imagenet classification with deep convolutional neural
  networks} {Imagenet classification with deep convolutional neural
  networks}.{\BBCQ}
\newblock
\BIn{} \APACrefbtitle {Advances in neural information processing systems}
  {Advances in neural information processing systems}\ (\BPGS\ 1097--1105).
\PrintBackRefs{\CurrentBib}

\bibitem [\protect \citeauthoryear {%
B.~Liu%
, Tang%
, Huang%
\BCBL {}\ \BBA {} Lu%
}{%
B.~Liu%
\ \protect \BOthers {.}}{%
{\protect \APACyear {2020}}%
}]{%
liu2020deep}
\APACinsertmetastar {%
liu2020deep}%
\begin{APACrefauthors}%
Liu, B.%
, Tang, J.%
, Huang, H.%
\BCBL {}\ \BBA {} Lu, X\BHBI Y.%
\end{APACrefauthors}%
\unskip\
\newblock
\APACrefYearMonthDay{2020}{}{}.
\newblock
{\BBOQ}\APACrefatitle {Deep learning methods for super-resolution
  reconstruction of turbulent flows} {Deep learning methods for
  super-resolution reconstruction of turbulent flows}.{\BBCQ}
\newblock
\APACjournalVolNumPages{Physics of Fluids}{32}{2}{025105}.
\PrintBackRefs{\CurrentBib}

\bibitem [\protect \citeauthoryear {%
Y.~Liu%
\ \protect \BOthers {.}}{%
Y.~Liu%
\ \protect \BOthers {.}}{%
{\protect \APACyear {2016}}%
}]{%
liu2016application}
\APACinsertmetastar {%
liu2016application}%
\begin{APACrefauthors}%
Liu, Y.%
, Racah, E.%
, Correa, J.%
, Khosrowshahi, A.%
, Lavers, D.%
, Kunkel, K.%
\BDBL {}others%
\end{APACrefauthors}%
\unskip\
\newblock
\APACrefYearMonthDay{2016}{}{}.
\newblock
{\BBOQ}\APACrefatitle {Application of deep convolutional neural networks for
  detecting extreme weather in climate datasets} {Application of deep
  convolutional neural networks for detecting extreme weather in climate
  datasets}.{\BBCQ}
\newblock
\APACjournalVolNumPages{arXiv preprint arXiv:1605.01156}{}{}{}.
\PrintBackRefs{\CurrentBib}

\bibitem [\protect \citeauthoryear {%
Lorenz%
}{%
Lorenz%
}{%
{\protect \APACyear {1956}}%
}]{%
lorenz1956empirical}
\APACinsertmetastar {%
lorenz1956empirical}%
\begin{APACrefauthors}%
Lorenz, E\BPBI N.%
\end{APACrefauthors}%
\unskip\
\newblock
\APACrefYearMonthDay{1956}{}{}.
\newblock
{\BBOQ}\APACrefatitle {Empirical orthogonal functions and statistical weather
  prediction} {Empirical orthogonal functions and statistical weather
  prediction}.{\BBCQ}
\newblock

\PrintBackRefs{\CurrentBib}

\bibitem [\protect \citeauthoryear {%
Lorenz%
}{%
Lorenz%
}{%
{\protect \APACyear {1963}}%
}]{%
lorenz1963deterministic}
\APACinsertmetastar {%
lorenz1963deterministic}%
\begin{APACrefauthors}%
Lorenz, E\BPBI N.%
\end{APACrefauthors}%
\unskip\
\newblock
\APACrefYearMonthDay{1963}{}{}.
\newblock
{\BBOQ}\APACrefatitle {Deterministic nonperiodic flow} {Deterministic
  nonperiodic flow}.{\BBCQ}
\newblock
\APACjournalVolNumPages{Journal of the atmospheric sciences}{20}{2}{130--141}.
\PrintBackRefs{\CurrentBib}

\bibitem [\protect \citeauthoryear {%
Lorenz%
}{%
Lorenz%
}{%
{\protect \APACyear {1996}}%
}]{%
lorenz1996predictability}
\APACinsertmetastar {%
lorenz1996predictability}%
\begin{APACrefauthors}%
Lorenz, E\BPBI N.%
\end{APACrefauthors}%
\unskip\
\newblock
\APACrefYearMonthDay{1996}{}{}.
\newblock
{\BBOQ}\APACrefatitle {Predictability: A problem partly solved}
  {Predictability: A problem partly solved}.{\BBCQ}
\newblock
\BIn{} \APACrefbtitle {Proc. Seminar on predictability} {Proc. seminar on
  predictability}\ (\BVOL~1).
\PrintBackRefs{\CurrentBib}

\bibitem [\protect \citeauthoryear {%
Majda%
, Timofeyev%
\BCBL {}\ \BBA {} Eijnden%
}{%
Majda%
\ \protect \BOthers {.}}{%
{\protect \APACyear {1999}}%
}]{%
majda1999models}
\APACinsertmetastar {%
majda1999models}%
\begin{APACrefauthors}%
Majda, A\BPBI J.%
, Timofeyev, I.%
\BCBL {}\ \BBA {} Eijnden, E\BPBI V.%
\end{APACrefauthors}%
\unskip\
\newblock
\APACrefYearMonthDay{1999}{}{}.
\newblock
{\BBOQ}\APACrefatitle {Models for stochastic climate prediction} {Models for
  stochastic climate prediction}.{\BBCQ}
\newblock
\APACjournalVolNumPages{Proceedings of the National Academy of
  Sciences}{96}{26}{14687--14691}.
\PrintBackRefs{\CurrentBib}

\bibitem [\protect \citeauthoryear {%
Maulik%
, San%
, Rasheed%
\BCBL {}\ \BBA {} Vedula%
}{%
Maulik%
\ \protect \BOthers {.}}{%
{\protect \APACyear {2019}}%
}]{%
maulik2019subgrid}
\APACinsertmetastar {%
maulik2019subgrid}%
\begin{APACrefauthors}%
Maulik, R.%
, San, O.%
, Rasheed, A.%
\BCBL {}\ \BBA {} Vedula, P.%
\end{APACrefauthors}%
\unskip\
\newblock
\APACrefYearMonthDay{2019}{}{}.
\newblock
{\BBOQ}\APACrefatitle {Subgrid modelling for two-dimensional turbulence using
  neural networks} {Subgrid modelling for two-dimensional turbulence using
  neural networks}.{\BBCQ}
\newblock
\APACjournalVolNumPages{Journal of Fluid Mechanics}{858}{}{122--144}.
\PrintBackRefs{\CurrentBib}

\bibitem [\protect \citeauthoryear {%
McGovern%
\ \protect \BOthers {.}}{%
McGovern%
\ \protect \BOthers {.}}{%
{\protect \APACyear {2019}}%
}]{%
mcgovern2019making}
\APACinsertmetastar {%
mcgovern2019making}%
\begin{APACrefauthors}%
McGovern, A.%
, Lagerquist, R.%
, Gagne, D\BPBI J.%
, Jergensen, G\BPBI E.%
, Elmore, K\BPBI L.%
, Homeyer, C\BPBI R.%
\BCBL {}\ \BBA {} Smith, T.%
\end{APACrefauthors}%
\unskip\
\newblock
\APACrefYearMonthDay{2019}{}{}.
\newblock
{\BBOQ}\APACrefatitle {Making the black box more transparent: Understanding the
  physical implications of machine learning} {Making the black box more
  transparent: Understanding the physical implications of machine
  learning}.{\BBCQ}
\newblock
\APACjournalVolNumPages{Bulletin of the American Meteorological
  Society}{100}{11}{2175--2199}.
\PrintBackRefs{\CurrentBib}

\bibitem [\protect \citeauthoryear {%
Mukhin%
\ \protect \BOthers {.}}{%
Mukhin%
\ \protect \BOthers {.}}{%
{\protect \APACyear {2015}}%
}]{%
mukhin2015}
\APACinsertmetastar {%
mukhin2015}%
\begin{APACrefauthors}%
Mukhin, D.%
, Kondrashov, D.%
, Loskutov, E.%
, Gavrilov, A.%
, Feigin, A.%
\BCBL {}\ \BBA {} Ghil, M.%
\end{APACrefauthors}%
\unskip\
\newblock
\APACrefYearMonthDay{2015}{}{}.
\newblock
{\BBOQ}\APACrefatitle {{Predicting critical transitions in ENSO models. Part
  II: Spatially dependent models}} {{Predicting critical transitions in ENSO
  models. Part II: Spatially dependent models}}.{\BBCQ}
\newblock
\APACjournalVolNumPages{Journal of Climate}{28}{5}{1962--1976}.
\PrintBackRefs{\CurrentBib}

\bibitem [\protect \citeauthoryear {%
Nadiga%
}{%
Nadiga%
}{%
{\protect \APACyear {2021}}%
}]{%
Nadiga2021}
\APACinsertmetastar {%
Nadiga2021}%
\begin{APACrefauthors}%
Nadiga, B.%
\end{APACrefauthors}%
\unskip\
\newblock
\APACrefYearMonthDay{2021}{}{}.
\newblock
{\BBOQ}\APACrefatitle {Reservoir Computing as a Tool for Climate Predictability
  Studies} {Reservoir computing as a tool for climate predictability
  studies}.{\BBCQ}
\newblock
\APACjournalVolNumPages{Journal of Advances in Modeling Earth
  Systems}{}{}{e2020MS002290}.
\newblock
\begin{APACrefDOI} \doi{https://doi.org/10.1029/2020MS002290} \end{APACrefDOI}
\PrintBackRefs{\CurrentBib}

\bibitem [\protect \citeauthoryear {%
Nielsen%
}{%
Nielsen%
}{%
{\protect \APACyear {2015}}%
}]{%
nielsen2015neural}
\APACinsertmetastar {%
nielsen2015neural}%
\begin{APACrefauthors}%
Nielsen, M\BPBI A.%
\end{APACrefauthors}%
\unskip\
\newblock
\APACrefYear{2015}.
\newblock
\APACrefbtitle {Neural networks and deep learning} {Neural networks and deep
  learning}\ (\BVOL\ 2018).
\newblock
\APACaddressPublisher{}{Determination press San Francisco, CA, USA}.
\PrintBackRefs{\CurrentBib}

\bibitem [\protect \citeauthoryear {%
Ollinaho%
\ \protect \BOthers {.}}{%
Ollinaho%
\ \protect \BOthers {.}}{%
{\protect \APACyear {2017}}%
}]{%
ollinaho2017towards}
\APACinsertmetastar {%
ollinaho2017towards}%
\begin{APACrefauthors}%
Ollinaho, P.%
, Lock, S\BHBI J.%
, Leutbecher, M.%
, Bechtold, P.%
, Beljaars, A.%
, Bozzo, A.%
\BDBL {}Sandu, I.%
\end{APACrefauthors}%
\unskip\
\newblock
\APACrefYearMonthDay{2017}{}{}.
\newblock
{\BBOQ}\APACrefatitle {Towards process-level representation of model
  uncertainties: stochastically perturbed parametrizations in the {ECMWF}
  ensemble} {Towards process-level representation of model uncertainties:
  stochastically perturbed parametrizations in the {ECMWF} ensemble}.{\BBCQ}
\newblock
\APACjournalVolNumPages{Quarterly Journal of the Royal Meteorological
  Society}{143}{702}{408--422}.
\PrintBackRefs{\CurrentBib}

\bibitem [\protect \citeauthoryear {%
Pathak%
, Hunt%
, Girvan%
, Lu%
\BCBL {}\ \BBA {} Ott%
}{%
Pathak%
\ \protect \BOthers {.}}{%
{\protect \APACyear {2018}}%
}]{%
RC}
\APACinsertmetastar {%
RC}%
\begin{APACrefauthors}%
Pathak, J.%
, Hunt, B.%
, Girvan, M.%
, Lu, Z.%
\BCBL {}\ \BBA {} Ott, E.%
\end{APACrefauthors}%
\unskip\
\newblock
\APACrefYearMonthDay{2018}{Jan}{}.
\newblock
{\BBOQ}\APACrefatitle {{Model-Free Prediction of Large Spatiotemporally Chaotic
  Systems from Data: A Reservoir Computing Approach}} {{Model-Free Prediction
  of Large Spatiotemporally Chaotic Systems from Data: A Reservoir Computing
  Approach}}.{\BBCQ}
\newblock
\APACjournalVolNumPages{Phys. Rev. Lett.}{120}{}{024102}.
\newblock
\begin{APACrefDOI} \doi{10.1103/PhysRevLett.120.024102} \end{APACrefDOI}
\PrintBackRefs{\CurrentBib}

\bibitem [\protect \citeauthoryear {%
Pawar%
, Ahmed%
, San%
\BCBL {}\ \BBA {} Rasheed%
}{%
Pawar%
\ \protect \BOthers {.}}{%
{\protect \APACyear {2020}}%
}]{%
pawar2020data}
\APACinsertmetastar {%
pawar2020data}%
\begin{APACrefauthors}%
Pawar, S.%
, Ahmed, S\BPBI E.%
, San, O.%
\BCBL {}\ \BBA {} Rasheed, A.%
\end{APACrefauthors}%
\unskip\
\newblock
\APACrefYearMonthDay{2020}{}{}.
\newblock
{\BBOQ}\APACrefatitle {Data-driven recovery of hidden physics in reduced order
  modeling of fluid flows} {Data-driven recovery of hidden physics in reduced
  order modeling of fluid flows}.{\BBCQ}
\newblock
\APACjournalVolNumPages{Physics of Fluids}{32}{3}{036602}.
\PrintBackRefs{\CurrentBib}

\bibitem [\protect \citeauthoryear {%
Penland%
}{%
Penland%
}{%
{\protect \APACyear {1989}}%
}]{%
Penland_MWR89}
\APACinsertmetastar {%
Penland_MWR89}%
\begin{APACrefauthors}%
Penland, C.%
\end{APACrefauthors}%
\unskip\
\newblock
\APACrefYearMonthDay{1989}{}{}.
\newblock
{\BBOQ}\APACrefatitle {Random forcing and forecasting using principal
  oscillation pattern analysis} {Random forcing and forecasting using principal
  oscillation pattern analysis}.{\BBCQ}
\newblock
\APACjournalVolNumPages{Monthly Weather Review}{117}{}{2165-2185}.
\PrintBackRefs{\CurrentBib}

\bibitem [\protect \citeauthoryear {%
Penland%
}{%
Penland%
}{%
{\protect \APACyear {1996}}%
}]{%
Penland_PD96}
\APACinsertmetastar {%
Penland_PD96}%
\begin{APACrefauthors}%
Penland, C.%
\end{APACrefauthors}%
\unskip\
\newblock
\APACrefYearMonthDay{1996}{}{}.
\newblock
{\BBOQ}\APACrefatitle {A stochastic model of {I}ndo-{P}acific sea surface
  temperature anomalies} {A stochastic model of {I}ndo-{P}acific sea surface
  temperature anomalies}.{\BBCQ}
\newblock
\APACjournalVolNumPages{Physica D}{98}{}{534-558}.
\PrintBackRefs{\CurrentBib}

\bibitem [\protect \citeauthoryear {%
Penland%
\ \BBA {} Sardeshmukh%
}{%
Penland%
\ \BBA {} Sardeshmukh%
}{%
{\protect \APACyear {1995}}%
}]{%
PenlandSardeshmukh_JCL95}
\APACinsertmetastar {%
PenlandSardeshmukh_JCL95}%
\begin{APACrefauthors}%
Penland, C.%
\BCBT {}\ \BBA {} Sardeshmukh, P\BPBI D.%
\end{APACrefauthors}%
\unskip\
\newblock
\APACrefYearMonthDay{1995}{}{}.
\newblock
{\BBOQ}\APACrefatitle {The optimal growth of tropical sea surface temperature
  anomalies} {The optimal growth of tropical sea surface temperature
  anomalies}.{\BBCQ}
\newblock
\APACjournalVolNumPages{Journal of {C}limate}{8}{}{1999-2024}.
\PrintBackRefs{\CurrentBib}

\bibitem [\protect \citeauthoryear {%
Portwood%
, Nadiga%
, Saenz%
\BCBL {}\ \BBA {} Livescu%
}{%
Portwood%
\ \protect \BOthers {.}}{%
{\protect \APACyear {2021}}%
}]{%
portwood2021interpreting}
\APACinsertmetastar {%
portwood2021interpreting}%
\begin{APACrefauthors}%
Portwood, G\BPBI D.%
, Nadiga, B\BPBI T.%
, Saenz, J\BPBI A.%
\BCBL {}\ \BBA {} Livescu, D.%
\end{APACrefauthors}%
\unskip\
\newblock
\APACrefYearMonthDay{2021}{}{}.
\newblock
{\BBOQ}\APACrefatitle {Interpreting neural network models of residual scalar
  flux} {Interpreting neural network models of residual scalar flux}.{\BBCQ}
\newblock
\APACjournalVolNumPages{Journal of Fluid Mechanics}{907}{}{}.
\PrintBackRefs{\CurrentBib}

\bibitem [\protect \citeauthoryear {%
Pyle%
, Jovanovic%
, Subramanian%
, Palem%
\BCBL {}\ \BBA {} Patel%
}{%
Pyle%
\ \protect \BOthers {.}}{%
{\protect \APACyear {2021}}%
}]{%
pyle2021domain}
\APACinsertmetastar {%
pyle2021domain}%
\begin{APACrefauthors}%
Pyle, R.%
, Jovanovic, N.%
, Subramanian, D.%
, Palem, K\BPBI V.%
\BCBL {}\ \BBA {} Patel, A\BPBI B.%
\end{APACrefauthors}%
\unskip\
\newblock
\APACrefYearMonthDay{2021}{}{}.
\newblock
{\BBOQ}\APACrefatitle {Domain-driven models yield better predictions at lower
  cost than reservoir computers in Lorenz systems} {Domain-driven models yield
  better predictions at lower cost than reservoir computers in lorenz
  systems}.{\BBCQ}
\newblock
\APACjournalVolNumPages{Philosophical Transactions of the Royal Society
  A}{379}{2194}{20200246}.
\PrintBackRefs{\CurrentBib}

\bibitem [\protect \citeauthoryear {%
Rahman%
, San%
, Rasheed%
\BCBL {}\ \protect \BOthers {.}}{%
Rahman%
\ \protect \BOthers {.}}{%
{\protect \APACyear {2018}}%
}]{%
rahman2018hybrid}
\APACinsertmetastar {%
rahman2018hybrid}%
\begin{APACrefauthors}%
Rahman, S.%
, San, O.%
, Rasheed, A.%
\BCBL {}\ \BOthersPeriod {.}\end{APACrefauthors}%
\unskip\
\newblock
\APACrefYearMonthDay{2018}{}{}.
\newblock
{\BBOQ}\APACrefatitle {A hybrid approach for model order reduction of
  barotropic quasi-geostrophic turbulence} {A hybrid approach for model order
  reduction of barotropic quasi-geostrophic turbulence}.{\BBCQ}
\newblock
\APACjournalVolNumPages{Fluids}{3}{4}{86}.
\PrintBackRefs{\CurrentBib}

\bibitem [\protect \citeauthoryear {%
Rasp%
, Pritchard%
\BCBL {}\ \BBA {} Gentine%
}{%
Rasp%
\ \protect \BOthers {.}}{%
{\protect \APACyear {2018}}%
}]{%
rasp2018deep}
\APACinsertmetastar {%
rasp2018deep}%
\begin{APACrefauthors}%
Rasp, S.%
, Pritchard, M\BPBI S.%
\BCBL {}\ \BBA {} Gentine, P.%
\end{APACrefauthors}%
\unskip\
\newblock
\APACrefYearMonthDay{2018}{}{}.
\newblock
{\BBOQ}\APACrefatitle {Deep learning to represent subgrid processes in climate
  models} {Deep learning to represent subgrid processes in climate
  models}.{\BBCQ}
\newblock
\APACjournalVolNumPages{Proceedings of the National Academy of
  Sciences}{115}{39}{9684--9689}.
\PrintBackRefs{\CurrentBib}

\bibitem [\protect \citeauthoryear {%
Rasp%
\ \BBA {} Thuerey%
}{%
Rasp%
\ \BBA {} Thuerey%
}{%
{\protect \APACyear {2021}}%
}]{%
rasp2021data}
\APACinsertmetastar {%
rasp2021data}%
\begin{APACrefauthors}%
Rasp, S.%
\BCBT {}\ \BBA {} Thuerey, N.%
\end{APACrefauthors}%
\unskip\
\newblock
\APACrefYearMonthDay{2021}{}{}.
\newblock
{\BBOQ}\APACrefatitle {Data-Driven Medium-Range Weather Prediction With a
  Resnet Pretrained on Climate Simulations: A New Model for WeatherBench}
  {Data-driven medium-range weather prediction with a resnet pretrained on
  climate simulations: A new model for weatherbench}.{\BBCQ}
\newblock
\APACjournalVolNumPages{Journal of Advances in Modeling Earth
  Systems}{13}{2}{e2020MS002405}.
\PrintBackRefs{\CurrentBib}

\bibitem [\protect \citeauthoryear {%
Reichstein%
\ \protect \BOthers {.}}{%
Reichstein%
\ \protect \BOthers {.}}{%
{\protect \APACyear {2019}}%
}]{%
reichstein2019deep}
\APACinsertmetastar {%
reichstein2019deep}%
\begin{APACrefauthors}%
Reichstein, M.%
, Camps-Valls, G.%
, Stevens, B.%
, Jung, M.%
, Denzler, J.%
, Carvalhais, N.%
\BCBL {}\ \BOthersPeriod {.}\end{APACrefauthors}%
\unskip\
\newblock
\APACrefYearMonthDay{2019}{}{}.
\newblock
{\BBOQ}\APACrefatitle {Deep learning and process understanding for data-driven
  Earth system science} {Deep learning and process understanding for
  data-driven earth system science}.{\BBCQ}
\newblock
\APACjournalVolNumPages{Nature}{566}{7743}{195--204}.
\PrintBackRefs{\CurrentBib}

\bibitem [\protect \citeauthoryear {%
Rowley%
\ \BBA {} Dawson%
}{%
Rowley%
\ \BBA {} Dawson%
}{%
{\protect \APACyear {2017}}%
}]{%
rowley2017model}
\APACinsertmetastar {%
rowley2017model}%
\begin{APACrefauthors}%
Rowley, C\BPBI W.%
\BCBT {}\ \BBA {} Dawson, S\BPBI T.%
\end{APACrefauthors}%
\unskip\
\newblock
\APACrefYearMonthDay{2017}{}{}.
\newblock
{\BBOQ}\APACrefatitle {Model reduction for flow analysis and control} {Model
  reduction for flow analysis and control}.{\BBCQ}
\newblock
\APACjournalVolNumPages{Annual Review of Fluid Mechanics}{49}{}{387--417}.
\PrintBackRefs{\CurrentBib}

\bibitem [\protect \citeauthoryear {%
Ryzhov%
, Kondrashov%
, Agarwal%
\BCBL {}\ \BBA {} Berloff%
}{%
Ryzhov%
\ \protect \BOthers {.}}{%
{\protect \APACyear {2019}}%
}]{%
ryzhov2019data}
\APACinsertmetastar {%
ryzhov2019data}%
\begin{APACrefauthors}%
Ryzhov, E.%
, Kondrashov, D.%
, Agarwal, N.%
\BCBL {}\ \BBA {} Berloff, P.%
\end{APACrefauthors}%
\unskip\
\newblock
\APACrefYearMonthDay{2019}{}{}.
\newblock
{\BBOQ}\APACrefatitle {On data-driven augmentation of low-resolution ocean
  model dynamics} {On data-driven augmentation of low-resolution ocean model
  dynamics}.{\BBCQ}
\newblock
\APACjournalVolNumPages{Ocean Modelling}{142}{}{101464}.
\PrintBackRefs{\CurrentBib}

\bibitem [\protect \citeauthoryear {%
Ryzhov%
, Kondrashov%
, Agarwal%
, McWilliams%
\BCBL {}\ \BBA {} Berloff%
}{%
Ryzhov%
\ \protect \BOthers {.}}{%
{\protect \APACyear {2020}}%
}]{%
RKNB2020}
\APACinsertmetastar {%
RKNB2020}%
\begin{APACrefauthors}%
Ryzhov, E.%
, Kondrashov, D.%
, Agarwal, N.%
, McWilliams, J.%
\BCBL {}\ \BBA {} Berloff, P.%
\end{APACrefauthors}%
\unskip\
\newblock
\APACrefYearMonthDay{2020}{}{}.
\newblock
{\BBOQ}\APACrefatitle {On data-driven induction of the low-frequency
  variability in a coarse-resolution ocean model} {On data-driven induction of
  the low-frequency variability in a coarse-resolution ocean model}.{\BBCQ}
\newblock
\APACjournalVolNumPages{Ocean Modelling}{153}{}{101664}.
\PrintBackRefs{\CurrentBib}

\bibitem [\protect \citeauthoryear {%
Salman%
, Heryadi%
, Abdurahman%
\BCBL {}\ \BBA {} Suparta%
}{%
Salman%
\ \protect \BOthers {.}}{%
{\protect \APACyear {2018}}%
}]{%
salman2018weather}
\APACinsertmetastar {%
salman2018weather}%
\begin{APACrefauthors}%
Salman, A\BPBI G.%
, Heryadi, Y.%
, Abdurahman, E.%
\BCBL {}\ \BBA {} Suparta, W.%
\end{APACrefauthors}%
\unskip\
\newblock
\APACrefYearMonthDay{2018}{}{}.
\newblock
{\BBOQ}\APACrefatitle {Weather Forecasting Using Merged Long Short-Term Memory
  Model ({LSTM}) and Autoregressive Integrated Moving Average ({ARIMA}) Model}
  {Weather forecasting using merged long short-term memory model ({LSTM}) and
  autoregressive integrated moving average ({ARIMA}) model}.{\BBCQ}
\newblock
\APACjournalVolNumPages{Journal of Computer Science}{14}{7}{930--938}.
\PrintBackRefs{\CurrentBib}

\bibitem [\protect \citeauthoryear {%
Sardeshmukh%
, Penland%
\BCBL {}\ \BBA {} Newman%
}{%
Sardeshmukh%
\ \protect \BOthers {.}}{%
{\protect \APACyear {2001}}%
}]{%
sardeshmukh2001rossby}
\APACinsertmetastar {%
sardeshmukh2001rossby}%
\begin{APACrefauthors}%
Sardeshmukh, P.%
, Penland, C.%
\BCBL {}\ \BBA {} Newman, M.%
\end{APACrefauthors}%
\unskip\
\newblock
\APACrefYearMonthDay{2001}{}{}.
\newblock
{\BBOQ}\APACrefatitle {Rossby waves in a stochastically fluctuating medium}
  {Rossby waves in a stochastically fluctuating medium}.{\BBCQ}
\newblock
\BIn{} \APACrefbtitle {Stochastic climate models} {Stochastic climate models}\
  (\BPGS\ 369--384).
\newblock
\APACaddressPublisher{}{Springer}.
\PrintBackRefs{\CurrentBib}

\bibitem [\protect \citeauthoryear {%
Sardeshmukh%
, Penland%
\BCBL {}\ \BBA {} Newman%
}{%
Sardeshmukh%
\ \protect \BOthers {.}}{%
{\protect \APACyear {2003}}%
}]{%
sardeshmukh2003drifts}
\APACinsertmetastar {%
sardeshmukh2003drifts}%
\begin{APACrefauthors}%
Sardeshmukh, P.%
, Penland, C.%
\BCBL {}\ \BBA {} Newman, M.%
\end{APACrefauthors}%
\unskip\
\newblock
\APACrefYearMonthDay{2003}{}{}.
\newblock
{\BBOQ}\APACrefatitle {Drifts induced by multiplicative red noise with
  application to climate} {Drifts induced by multiplicative red noise with
  application to climate}.{\BBCQ}
\newblock
\APACjournalVolNumPages{EPL (Europhysics Letters)}{63}{4}{498}.
\PrintBackRefs{\CurrentBib}

\bibitem [\protect \citeauthoryear {%
Scher%
\ \BBA {} Messori%
}{%
Scher%
\ \BBA {} Messori%
}{%
{\protect \APACyear {2019}}%
}]{%
scher2019generalization}
\APACinsertmetastar {%
scher2019generalization}%
\begin{APACrefauthors}%
Scher, S.%
\BCBT {}\ \BBA {} Messori, G.%
\end{APACrefauthors}%
\unskip\
\newblock
\APACrefYearMonthDay{2019}{}{}.
\newblock
{\BBOQ}\APACrefatitle {Generalization properties of feed-forward neural
  networks trained on Lorenz systems} {Generalization properties of
  feed-forward neural networks trained on lorenz systems}.{\BBCQ}
\newblock
\APACjournalVolNumPages{Nonlinear processes in geophysics}{26}{4}{381--399}.
\PrintBackRefs{\CurrentBib}

\bibitem [\protect \citeauthoryear {%
Seiffert%
\ \BBA {} Von~Storch%
}{%
Seiffert%
\ \BBA {} Von~Storch%
}{%
{\protect \APACyear {2008}}%
}]{%
seiffert2008impact}
\APACinsertmetastar {%
seiffert2008impact}%
\begin{APACrefauthors}%
Seiffert, R.%
\BCBT {}\ \BBA {} Von~Storch, J\BHBI S.%
\end{APACrefauthors}%
\unskip\
\newblock
\APACrefYearMonthDay{2008}{}{}.
\newblock
{\BBOQ}\APACrefatitle {Impact of atmospheric small-scale fluctuations on
  climate sensitivity} {Impact of atmospheric small-scale fluctuations on
  climate sensitivity}.{\BBCQ}
\newblock
\APACjournalVolNumPages{Geophysical research letters}{35}{10}{}.
\PrintBackRefs{\CurrentBib}

\bibitem [\protect \citeauthoryear {%
Seiffert%
\ \BBA {} von Storch%
}{%
Seiffert%
\ \BBA {} von Storch%
}{%
{\protect \APACyear {2010}}%
}]{%
seiffert2010stochastic}
\APACinsertmetastar {%
seiffert2010stochastic}%
\begin{APACrefauthors}%
Seiffert, R.%
\BCBT {}\ \BBA {} von Storch, J\BHBI S.%
\end{APACrefauthors}%
\unskip\
\newblock
\APACrefYearMonthDay{2010}{}{}.
\newblock
{\BBOQ}\APACrefatitle {A stochastic analysis of the impact of small-scale
  fluctuations on the tropospheric temperature response to {CO2} doubling} {A
  stochastic analysis of the impact of small-scale fluctuations on the
  tropospheric temperature response to {CO2} doubling}.{\BBCQ}
\newblock
\APACjournalVolNumPages{Journal of Climate}{23}{9}{2307--2319}.
\PrintBackRefs{\CurrentBib}

\bibitem [\protect \citeauthoryear {%
Seleznev%
, Mukhin%
, Gavrilov%
, Loskutov%
\BCBL {}\ \BBA {} Feigin%
}{%
Seleznev%
\ \protect \BOthers {.}}{%
{\protect \APACyear {2019}}%
}]{%
Seleznev2019}
\APACinsertmetastar {%
Seleznev2019}%
\begin{APACrefauthors}%
Seleznev, A.%
, Mukhin, D.%
, Gavrilov, A.%
, Loskutov, E.%
\BCBL {}\ \BBA {} Feigin, A.%
\end{APACrefauthors}%
\unskip\
\newblock
\APACrefYearMonthDay{2019}{}{}.
\newblock
{\BBOQ}\APACrefatitle {Bayesian framework for simulation of dynamical systems
  from multidimensional data using recurrent neural network} {Bayesian
  framework for simulation of dynamical systems from multidimensional data
  using recurrent neural network}.{\BBCQ}
\newblock
\APACjournalVolNumPages{Chaos}{29}{12}{123115}.
\newblock
\begin{APACrefDOI} \doi{10.1063/1.5128372} \end{APACrefDOI}
\PrintBackRefs{\CurrentBib}

\bibitem [\protect \citeauthoryear {%
Sexton%
, Murphy%
, Collins%
\BCBL {}\ \BBA {} Webb%
}{%
Sexton%
\ \protect \BOthers {.}}{%
{\protect \APACyear {2012}}%
}]{%
sexton2012multivariate}
\APACinsertmetastar {%
sexton2012multivariate}%
\begin{APACrefauthors}%
Sexton, D\BPBI M.%
, Murphy, J\BPBI M.%
, Collins, M.%
\BCBL {}\ \BBA {} Webb, M\BPBI J.%
\end{APACrefauthors}%
\unskip\
\newblock
\APACrefYearMonthDay{2012}{}{}.
\newblock
{\BBOQ}\APACrefatitle {Multivariate probabilistic projections using imperfect
  climate models part I: outline of methodology} {Multivariate probabilistic
  projections using imperfect climate models part i: outline of
  methodology}.{\BBCQ}
\newblock
\APACjournalVolNumPages{Climate dynamics}{38}{11-12}{2513--2542}.
\PrintBackRefs{\CurrentBib}

\bibitem [\protect \citeauthoryear {%
Strounine%
, Kravtsov%
, Kondrashov%
\BCBL {}\ \BBA {} Ghil%
}{%
Strounine%
\ \protect \BOthers {.}}{%
{\protect \APACyear {2010}}%
}]{%
Strounine2010}
\APACinsertmetastar {%
Strounine2010}%
\begin{APACrefauthors}%
Strounine, K.%
, Kravtsov, S.%
, Kondrashov, D.%
\BCBL {}\ \BBA {} Ghil, M.%
\end{APACrefauthors}%
\unskip\
\newblock
\APACrefYearMonthDay{2010}{}{}.
\newblock
{\BBOQ}\APACrefatitle {Reduced models of atmospheric low-frequency variability:
  Parameter estimation and comparative performance} {Reduced models of
  atmospheric low-frequency variability: Parameter estimation and comparative
  performance}.{\BBCQ}
\newblock
\APACjournalVolNumPages{Physica D: Nonlinear Phenomena}{239}{3}{145-166}.
\newblock
\begin{APACrefDOI} \doi{https://doi.org/10.1016/j.physd.2009.10.013}
  \end{APACrefDOI}
\PrintBackRefs{\CurrentBib}

\bibitem [\protect \citeauthoryear {%
Sura%
}{%
Sura%
}{%
{\protect \APACyear {2011}}%
}]{%
sura2011general}
\APACinsertmetastar {%
sura2011general}%
\begin{APACrefauthors}%
Sura, P.%
\end{APACrefauthors}%
\unskip\
\newblock
\APACrefYearMonthDay{2011}{}{}.
\newblock
{\BBOQ}\APACrefatitle {A general perspective of extreme events in weather and
  climate} {A general perspective of extreme events in weather and
  climate}.{\BBCQ}
\newblock
\APACjournalVolNumPages{Atmospheric Research}{101}{1-2}{1--21}.
\PrintBackRefs{\CurrentBib}

\bibitem [\protect \citeauthoryear {%
Sura%
}{%
Sura%
}{%
{\protect \APACyear {2013}}%
}]{%
sura2013stochastic}
\APACinsertmetastar {%
sura2013stochastic}%
\begin{APACrefauthors}%
Sura, P.%
\end{APACrefauthors}%
\unskip\
\newblock
\APACrefYearMonthDay{2013}{}{}.
\newblock
{\BBOQ}\APACrefatitle {Stochastic models of climate extremes: Theory and
  observations} {Stochastic models of climate extremes: Theory and
  observations}.{\BBCQ}
\newblock
\BIn{} \APACrefbtitle {Extremes in a Changing Climate} {Extremes in a changing
  climate}\ (\BPGS\ 181--222).
\newblock
\APACaddressPublisher{}{Springer}.
\PrintBackRefs{\CurrentBib}

\bibitem [\protect \citeauthoryear {%
Sura%
, Newman%
, Penland%
\BCBL {}\ \BBA {} Sardeshmukh%
}{%
Sura%
\ \protect \BOthers {.}}{%
{\protect \APACyear {2005}}%
}]{%
sura2005multiplicative}
\APACinsertmetastar {%
sura2005multiplicative}%
\begin{APACrefauthors}%
Sura, P.%
, Newman, M.%
, Penland, C.%
\BCBL {}\ \BBA {} Sardeshmukh, P.%
\end{APACrefauthors}%
\unskip\
\newblock
\APACrefYearMonthDay{2005}{}{}.
\newblock
{\BBOQ}\APACrefatitle {Multiplicative noise and non-Gaussianity: A paradigm for
  atmospheric regimes?} {Multiplicative noise and non-gaussianity: A paradigm
  for atmospheric regimes?}{\BBCQ}
\newblock
\APACjournalVolNumPages{Journal of the atmospheric
  sciences}{62}{5}{1391--1409}.
\PrintBackRefs{\CurrentBib}

\bibitem [\protect \citeauthoryear {%
Vlachas%
, Byeon%
, Wan%
, Sapsis%
\BCBL {}\ \BBA {} Koumoutsakos%
}{%
Vlachas%
\ \protect \BOthers {.}}{%
{\protect \APACyear {2018}}%
}]{%
vlachas2018data}
\APACinsertmetastar {%
vlachas2018data}%
\begin{APACrefauthors}%
Vlachas, P\BPBI R.%
, Byeon, W.%
, Wan, Z\BPBI Y.%
, Sapsis, T\BPBI P.%
\BCBL {}\ \BBA {} Koumoutsakos, P.%
\end{APACrefauthors}%
\unskip\
\newblock
\APACrefYearMonthDay{2018}{}{}.
\newblock
{\BBOQ}\APACrefatitle {Data-driven forecasting of high-dimensional chaotic
  systems with long short-term memory networks} {Data-driven forecasting of
  high-dimensional chaotic systems with long short-term memory
  networks}.{\BBCQ}
\newblock
\APACjournalVolNumPages{Proceedings of the Royal Society A: Mathematical,
  Physical and Engineering Sciences}{474}{2213}{20170844}.
\PrintBackRefs{\CurrentBib}

\bibitem [\protect \citeauthoryear {%
Wang%
, Ripamonti%
\BCBL {}\ \BBA {} Hesthaven%
}{%
Wang%
\ \protect \BOthers {.}}{%
{\protect \APACyear {2020}}%
}]{%
WANG2020}
\APACinsertmetastar {%
WANG2020}%
\begin{APACrefauthors}%
Wang, Q.%
, Ripamonti, N.%
\BCBL {}\ \BBA {} Hesthaven, J\BPBI S.%
\end{APACrefauthors}%
\unskip\
\newblock
\APACrefYearMonthDay{2020}{}{}.
\newblock
{\BBOQ}\APACrefatitle {{Recurrent neural network closure of parametric
  POD-Galerkin reduced-order models based on the Mori-Zwanzig formalism}}
  {{Recurrent neural network closure of parametric POD-Galerkin reduced-order
  models based on the Mori-Zwanzig formalism}}.{\BBCQ}
\newblock
\APACjournalVolNumPages{Journal of Computational Physics}{410}{}{109402}.
\newblock
\begin{APACrefDOI} \doi{https://doi.org/10.1016/j.jcp.2020.109402}
  \end{APACrefDOI}
\PrintBackRefs{\CurrentBib}

\bibitem [\protect \citeauthoryear {%
Watson%
}{%
Watson%
}{%
{\protect \APACyear {2019}}%
}]{%
watson2019applying}
\APACinsertmetastar {%
watson2019applying}%
\begin{APACrefauthors}%
Watson, P\BPBI A.%
\end{APACrefauthors}%
\unskip\
\newblock
\APACrefYearMonthDay{2019}{}{}.
\newblock
{\BBOQ}\APACrefatitle {Applying machine learning to improve simulations of a
  chaotic dynamical system using empirical error correction} {Applying machine
  learning to improve simulations of a chaotic dynamical system using empirical
  error correction}.{\BBCQ}
\newblock
\APACjournalVolNumPages{Journal of advances in modeling earth
  systems}{11}{5}{1402--1417}.
\PrintBackRefs{\CurrentBib}

\bibitem [\protect \citeauthoryear {%
Weyn%
, Durran%
\BCBL {}\ \BBA {} Caruana%
}{%
Weyn%
\ \protect \BOthers {.}}{%
{\protect \APACyear {2020}}%
}]{%
weyn2020improving}
\APACinsertmetastar {%
weyn2020improving}%
\begin{APACrefauthors}%
Weyn, J\BPBI A.%
, Durran, D\BPBI R.%
\BCBL {}\ \BBA {} Caruana, R.%
\end{APACrefauthors}%
\unskip\
\newblock
\APACrefYearMonthDay{2020}{}{}.
\newblock
{\BBOQ}\APACrefatitle {Improving Data-Driven Global Weather Prediction Using
  Deep Convolutional Neural Networks on a Cubed Sphere} {Improving data-driven
  global weather prediction using deep convolutional neural networks on a cubed
  sphere}.{\BBCQ}
\newblock
\APACjournalVolNumPages{Journal of Advances in Modeling Earth
  Systems}{12}{9}{e2020MS002109}.
\PrintBackRefs{\CurrentBib}

\bibitem [\protect \citeauthoryear {%
Williams%
, Howe%
, Gregory%
, Smith%
\BCBL {}\ \BBA {} Joshi%
}{%
Williams%
\ \protect \BOthers {.}}{%
{\protect \APACyear {2016}}%
}]{%
williams2016improved}
\APACinsertmetastar {%
williams2016improved}%
\begin{APACrefauthors}%
Williams, P\BPBI D.%
, Howe, N\BPBI J.%
, Gregory, J\BPBI M.%
, Smith, R\BPBI S.%
\BCBL {}\ \BBA {} Joshi, M\BPBI M.%
\end{APACrefauthors}%
\unskip\
\newblock
\APACrefYearMonthDay{2016}{}{}.
\newblock
{\BBOQ}\APACrefatitle {Improved climate simulations through a stochastic
  parameterization of ocean eddies} {Improved climate simulations through a
  stochastic parameterization of ocean eddies}.{\BBCQ}
\newblock
\APACjournalVolNumPages{Journal of Climate}{29}{24}{8763--8781}.
\PrintBackRefs{\CurrentBib}

\bibitem [\protect \citeauthoryear {%
Williamson%
}{%
Williamson%
}{%
{\protect \APACyear {2015}}%
}]{%
williamson2015exploratory}
\APACinsertmetastar {%
williamson2015exploratory}%
\begin{APACrefauthors}%
Williamson, D.%
\end{APACrefauthors}%
\unskip\
\newblock
\APACrefYearMonthDay{2015}{}{}.
\newblock
{\BBOQ}\APACrefatitle {Exploratory ensemble designs for environmental models
  using k-extended Latin Hypercubes} {Exploratory ensemble designs for
  environmental models using k-extended latin hypercubes}.{\BBCQ}
\newblock
\APACjournalVolNumPages{Environmetrics}{26}{4}{268--283}.
\PrintBackRefs{\CurrentBib}

\bibitem [\protect \citeauthoryear {%
Q.~Zhang%
, Wang%
, Dong%
, Zhong%
\BCBL {}\ \BBA {} Sun%
}{%
Q.~Zhang%
\ \protect \BOthers {.}}{%
{\protect \APACyear {2017}}%
}]{%
zhang2017prediction}
\APACinsertmetastar {%
zhang2017prediction}%
\begin{APACrefauthors}%
Zhang, Q.%
, Wang, H.%
, Dong, J.%
, Zhong, G.%
\BCBL {}\ \BBA {} Sun, X.%
\end{APACrefauthors}%
\unskip\
\newblock
\APACrefYearMonthDay{2017}{}{}.
\newblock
{\BBOQ}\APACrefatitle {Prediction of sea surface temperature using long
  short-term memory} {Prediction of sea surface temperature using long
  short-term memory}.{\BBCQ}
\newblock
\APACjournalVolNumPages{IEEE Geoscience and Remote Sensing
  Letters}{14}{10}{1745--1749}.
\PrintBackRefs{\CurrentBib}

\bibitem [\protect \citeauthoryear {%
Y.~Zhang%
\ \BBA {} Held%
}{%
Y.~Zhang%
\ \BBA {} Held%
}{%
{\protect \APACyear {1999}}%
}]{%
zhang1999linear}
\APACinsertmetastar {%
zhang1999linear}%
\begin{APACrefauthors}%
Zhang, Y.%
\BCBT {}\ \BBA {} Held, I\BPBI M.%
\end{APACrefauthors}%
\unskip\
\newblock
\APACrefYearMonthDay{1999}{}{}.
\newblock
{\BBOQ}\APACrefatitle {A linear stochastic model of a {GCM}’s midlatitude
  storm tracks} {A linear stochastic model of a {GCM}’s midlatitude storm
  tracks}.{\BBCQ}
\newblock
\APACjournalVolNumPages{Journal of the atmospheric
  sciences}{56}{19}{3416--3435}.
\PrintBackRefs{\CurrentBib}

\end{thebibliography}

\end{document}